\documentclass[aas2,natbib209]{aastex}
\usepackage{amssymb}
\usepackage{amsmath}
\usepackage{fixltx2e}
\usepackage{float}
\usepackage{natbib}
\usepackage{color} 

\newcommand{\gtsimeq}{\raisebox{-0.6ex}{$\, \stackrel{\raisebox{-.2ex}%
{$\textstyle >$}}{\sim}\,$}}
\newcommand{\ltsimeq}{\raisebox{-0.6ex}{$\, \stackrel{\raisebox{-.2ex}%
{$\textstyle <$}}{\sim}\,$}}
\newcommand{\kp}{\mbox{${\rm K}^{+}$}}
\newcommand{\kz}{\mbox{${\rm K}^{0}$}}

\shorttitle{High-Temperature Ionization in Protoplanetary Disks}
\shortauthors{Desch and Turner}

\begin{document}

\title{High-Temperature Ionization in Protoplanetary Disks}

\author{Steven J.\ Desch}
\affil{School of Earth and Space Exploration, Arizona State University, PO Box 871404, Tempe, AZ 85287-1404}

\author{Neal J.\ Turner}
\affil{Jet Propulsion Laboratory, Mail Stop 169-506, California Institute of Technology, 4800 Oak Grove Drive, Pasadena, CA 91109}

\begin{abstract}
We calculate the abundances of electrons and ions in the hot
($\gtsimeq \, 500$~K), dusty parts of protoplanetary disks, treating
for the first time the effects of thermionic and ion emission from the
dust grains.  High-temperature ionization modeling has involved simply
assuming that alkali elements such as potassium occur as gas-phase
atoms and are collisionally ionized following the Saha equation.  We
show that the Saha equation often does not hold, because free charges
are produced by thermionic and ion emission and destroyed when they
stick to grain surfaces.  This means the ionization state depends not
on the first ionization potential of the alkali atoms, but rather on
the grains' work functions.  The charged species' abundances typically
rise abruptly above about 800~K, with little qualitative dependence on
the work function, gas density, or dust-to-gas mass ratio.  Applying
our results, we find that protoplanetary disks' dead zone, where high
diffusivities stifle magnetorotational turbulence, has its inner edge
located where the temperature exceeds a threshold value $\approx
1000$~K.  The threshold is set by ambipolar diffusion except at the
highest densities, where it is set by Ohmic resistivity.  We find that
the disk gas can be diffusively loaded onto the stellar magnetosphere
at temperatures below a similar threshold.  We investigate whether the
``short-circuit'' instability of current sheets can operate in disks
and find that it cannot, or works only in a narrow range of
conditions; it appears not to be the chondrule formation mechanism.
We also suggest that thermionic emission is important for determining
the rate of Ohmic heating in hot Jupiters.
\end{abstract}



\section{Introduction}

Protoplanetary disks' evolution is governed by the extraction of
orbital angular momentum needed for accretion to proceed.  Magnetic
forces are among the prime candidates to drive the extraction.
Modeling the disks therefore requires knowing how readily the magnetic
fields diffuse through the material.  Hence the ionization state of
the gas is key.

In particular, the magnetorotational instability (MRI) is suppressed
unless the electron fraction is $x_{\rm e} = n_{\rm e} / n_{\rm H2} \,
\gtsimeq \, 10^{-13}$ (Jin 1996; Gammie 1996), where $n_{\rm H2}$ is
the number density of ${\rm H}_{2}$ molecules.  Launching disk winds
requires $x_{\rm e} \, \gtsimeq \, 10^{-11} - 10^{-10}$ (Bai \& Stone
2013).  The gas ionization likewise bears on the existence of
lightning (Desch \& Cuzzi 2000) and current sheets (McNally et
al.\ 2013), two proposed mechanisms for transiently heating and
melting chondrules (millimeter-sized inclusions in chondritic
meteorites) in the solar nebula.  Assessing the extent of the MRI and
the heating of chondrules requires knowing the ionization rates,
especially in the disks' denser midplane regions.

Ionization in the cooler parts of disks, where temperatures $T \,
\ltsimeq \, 500$~K, has been studied in detail.  Several non-thermal
processes are important at these low temperatures, with ionization
rates per volume often expressed as $\zeta \, n_{\rm H2}$.
Far-ultraviolet (FUV) photons from the central star can ionize gas at
a rate $\zeta_{\rm UV} \sim 10^{-8} \, {\rm s}^{-1}$ at 1~AU, but are
attenuated by column densities $\ltsimeq \, 0.1 \ {\rm g} \, {\rm
  cm}^{-3}$ (Perez-Becker \& Chiang 2011).  Similarly, X-rays from the
star can ionize at rates $\zeta_{\rm X} \sim 10^{-11} \, {\rm s}^{-1}$
at 1~AU, but are attenuated by column densities $\ltsimeq \, 10 \ {\rm
  g} \, {\rm cm}^{-3}$ (Glassgold et al.\ 1997; Igea \& Glassgold
1999; Ercolano \& Glassgold 2013).  Galactic cosmic rays, if not
shielded by stellar winds (Cleeves et al.\ 2013a), can ionize at rates
$\zeta_{\rm GCR} \sim 10^{-17} \, {\rm s}^{-1}$, but are attenuated by
column densities $\sim 100 \ {\rm g} \, {\rm cm}^{-3}$ (Umebayashi \&
Nakano 1981).  In the innermost few AU of a protoplanetary disk, the
expected column densities are $\gg 10^3 \, {\rm g} \, {\rm cm}^{-2}$,
so that the dominant non-thermal ionization sources are $\beta$
particles from radioactive decays of species like ${}^{40}{\rm K}$,
leading to $\zeta_{\rm rad} \sim 10^{-21} \, {\rm s}^{-1}$ (Umebayashi
\& Nakano 1981); or ${}^{26}{\rm Al}$ (if present), leading to
$\zeta_{\rm rad} \, \ltsimeq \, 10^{-18} \, {\rm s}^{-1}$ (Cleeves et
al.\ 2013b).  In the disks' cold midplane regions, this is the
expected ionization rate, and it is quite low.

In chemical equilibrium in the dense midplane gas, the electron
density is found by balancing the ionization rate against adsorption,
and subsequent recombination, of electrons on grain surfaces.
Assuming monodisperse grains with radius $a_{\rm gr} = 1 \,\mu{\rm
  m}$, internal density $\rho_{\rm s} = 3 \, {\rm g} \, {\rm
  cm}^{-3}$, and dust-to-gas mass ratio $\rho_{\rm gr} / \rho_{\rm g}
= 0.01$, the equation describing this balance is
\begin{equation}
\zeta \, n_{{\rm H}_{2}} = n_{\rm gr} \, \pi a_{\rm gr}^2 \, n_{\rm e} \,
 \left( \frac{ 8 k T }{ \pi m_{\rm e} } \right)^{1/2} \, S_{\rm e}, 
\end{equation}
(Sano et al.\ 2000), where $n_{\rm gr} = (\rho_{\rm gr} / \rho_{\rm
  g}) \, (1.4 \, n_{{\rm H}_{2}} \, m_{{\rm H}_{2}}) \, / (4\pi
\rho_{\rm s} a_{\rm gr}^{3} / 3)$ is the number density of grains,
$S_{\rm e}$ the sticking coefficient of electrons colliding with
grains, and other variables have their usual meanings.  From this we
find $n_{\rm e} \sim 10^{-3} \, {\rm cm}^{-3}$ (assuming $\zeta =
10^{-18} \, {\rm s}^{-1}$, $T = 300 \, {\rm K}$ and $S_{\rm e} =
0.6$).  For a typical midplane gas density inside a few AU, $n_{{\rm
    H}_{2}} \, \gtsimeq \, 10^{14} \, {\rm cm}^{-3}$ (Weidenschilling
1977), $x_{\rm e} \, \ltsimeq \, 10^{-17}$, a value too low to be of
significance to the MRI.  Recognition of this fact led Gammie (1996)
to suggest that protoplanetary disks' interiors, especially the parts
within a few AU of their stars, are ``dead zones'' where the MRI
cannot act.

Jin (1996) and Gammie (1996) were the first to recognize that very
close to the central star, protoplanetary disks may be hot enough to
sustain thermal ionizations, recoupling the magnetic field to the gas.
Specifically, they pointed out that collisional ionization of
gas-phase potassium atoms at temperatures $\gtsimeq \, 1000 \, {\rm
  K}$ would generate free electrons and ions.  The threshold
temperature of 1000~K is based on calculating the abundances of ${\rm
  K}^{+}$ ions and electrons using the Saha equation:
\begin{equation}
\frac{ n_{{\rm K}^{+}}   }{ n_{{\rm K}^{0}} }
= \frac{ g_{+} }{ g_{0} } \,
\frac{2}{ n_{\rm e} } \, \left( \frac{ 2\pi m_{\rm e} k T }{ h^2 } \right)^{3/2} \, 
\exp \left( -\frac{\rm IP}{k T} \right).
\end{equation}
Here $g_{+} = 1$ and $g_{0} = 2$ are the statistical weights of the
(ground states) of $\kp$ and $\kz$, $k$ the Boltzmann constant, $h$
Planck's constant, $m_{\rm e}$ the electron mass, and ${\rm IP} = 4.34
\, {\rm eV}$ the first ionization potential energy of potassium.
Assuming $n_{{\rm K}^{+}} = n_{\rm e}$ and that the abundance of K
relative to ${\rm H}_{2}$ in a solar-composition gas is $x_{\rm K} =
3.0 \times 10^{-7}$ (Lodders 2003), half of the K atoms are ionized
when $T = 1715 \, {\rm K}$, for a density $n_{{\rm H}_{2}} = 10^{14}
\, {\rm cm}^{-3}$ likely to occur near the midplane.  More
significantly for the MRI, $x_{\rm e} = 10^{-13}$ when $T = 865 \,
{\rm K}$, meaning substantial ionization would be expected if
potassium were in the gas phase at this temperature.  According to
Lodders (2003), the 50\% condensation temperature of potassium is
1006~K (at $P = 10^{-4} \, {\rm bar}$, or $n_{{\rm H}_{2}} = 6 \times
10^{14} \, {\rm cm}^{-3}$; the pressure dependence is weak).  So it is
fairer to say that above 1000~K the Saha equation predicts significant
thermal ionization of potassium, and for temperatures between about
850 and 1000~K, thermal ionization may or may not lead to significant
ionization.

Other species contribute less strongly to thermal ionization than
potassium.  Because of the exponential dependence on the first
ionization potential (${\rm IP}$), one might expect other alkali atoms
to contribute to thermal ionization.  The first ionization potentials
of Li, Na, K, Rb, and Cs are, respectively, 5.39, 5.14, 4.34, 4.18,
and 3.89~eV.  These elements' abundances relative to ${\rm H}_{2}$
are, respectively, $4.57 \times 10^{-9}$, $4.73 \times 10^{-6}$, $3.04
\times 10^{-7}$, $5.51 \times 10^{-10}$, and $3.02 \times 10^{-11}$
(Lodders 2003).  The first ionization potentials (FIP) and
cosmochemical abundances of many elements are depicted in Figure~1.
Repeating the analysis above, the temperature at which each alkali
element contributes enough electrons to yield $x_{\rm e} = 10^{-13}$
(at $n_{{\rm H}_{2}} = 10^{14} \, {\rm cm}^{-3}$) is 1149, 975, 865,
932, and 852~K, respectively.  Considering that cesium is in the gas
phase above 800~K whereas potassium is not (Lodders 2003), one might
expect Cs to be the dominant contributor of ions and electrons at
temperatures between 850 and 1000~K; however, the Cs is $10^{4}$ times
less abundant, so that even if all Cs were ionized and only 0.1\% of
all K atoms were in the gas phase, potassium would still dominate over
cesium.  Thermal ionization of potassium is likely to be more
important than that of other alkalis, and potassium is in any case an
excellent proxy for other alkalis.  Throughout this paper we will
consider the case of potassium to represent the contributions of all
the alkali elements to thermal ionization.
%-------------------------------------------------------------------------------
%
% FIGURE 1
%
\begin{figure}[H]
\begin{center} 
\includegraphics[width=0.80\paperwidth]{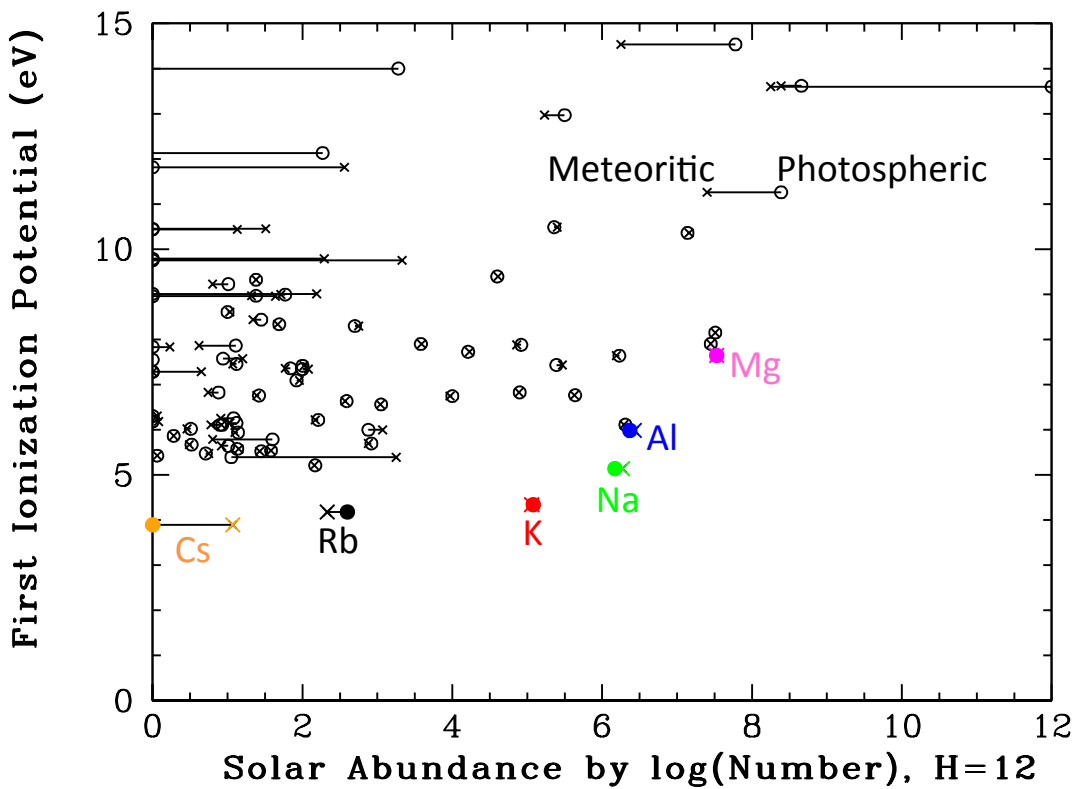}
\caption{Cosmochemical abundances of various elements and their first
  ionization potentials (FIPs).  Abundances (on a scale where the
  number of H atoms is $10^{12}$) are derived by Asplund et
  al.\ (2005) from analyses of carbonaceous chondrites
  (``meteoritic'') or from spectrometry of the Sun (``photospheric'').
  Producing free electrons by thermal ionization requires the presence
  of elements with high abundance and low FIP, i.e., those elements
  toward the plot's lower right corner.  The exponential sensitivity
  of the ionization fraction to FIP means potassium (K) is most
  favored to provide free electrons by thermal ionization, with Na,
  Rb, and Cs potentially also important---if these elements occur in
  the gas phase rather than locked up in minerals. }
\end{center}
\label{fig:thermalionization}
\end{figure}
%
%-------------------------------------------------------------------------------

A criticism of this thermal ionization picture is that the Saha
equation is valid only in local thermodynamic equilibrium, or LTE
(Desch 1998).  It relies on the collisional ionization of the
gas-phase K atoms, e.g., ${\rm H}_{2} + {\rm K}^{0} \rightarrow {\rm
  H}_{2} + {\rm K}^{+} + {\rm e}^{-}$, being in detailed balance with
its inverse reaction, three-body gas-phase recombination, ${\rm H}_{2}
+ {\rm K}^{+} + {\rm e}^{-} \rightarrow {\rm H}_{2} + {\rm K}^{0}$.
But under conditions typical for protoplanetary disks, electrons are
more likely to recombine with ions on the surfaces of grains.  Desch
(1998) argued that the electron density must be set by balancing
collisional ionizations against collisions of electrons with dust
grains.  Once adsorbed on the grain surfaces, electrons and ions
quantum tunnel from one adsorption site to the next, wandering across
the surface until they find each other and recombine.  The resulting
neutral K atoms can then escape the grain.  Dust grain
surface-catalyzed recombinations are much faster than gas-phase
recombinations, as the frequency of colliding with dust grains is
$\sim n_{\rm gr} \, \pi a_{\rm gr}^2 \, (8 k T / \pi m_{\rm e})^{1/2}
\, S_{\rm e}$ $\sim 0.1 \, {\rm s}^{-1}$ (at $T = 1000$ K and $n_{{\rm
    H}_{2}} = 10^{14} \, {\rm cm}^{-3}$, and assuming grains with
radius $1 \, \mu{\rm m}$ and a dust-to-gas mass ratio 0.01), whereas
the frequency of recombining with an ion is $\sim k_{-2} n_{{\rm
    H}_{2}} \, n_{{\rm K}^{+}} \, \ltsimeq \, 10^{-11} \, {\rm
  s}^{-1}$, where $k_{-2}$ is the recombination coefficient (see
below), and we have assumed the ion density matches the electron
density, and $x_{\rm e} = 10^{-11}$.  Not until $x_{\rm e} \, \gtsimeq
\, 10^{-3}$ do gas-phase recombinations dominate over grain-catalyzed
recombinations.  The dominance of grain effects invalidates the use of
the Saha equation for potassium as a means of calculating the electron
density.

The dominance of grain processes also points to an additional
ionization source, not previously considered in the context of
protoplanetary disks.  The process that is the inverse of electrons
and ions colliding with grains, resulting in neutral atoms leaving the
grain, is the emission of electrons or ions following collisions of
neutral atoms with grains.  These processes are more commonly called
thermionic emission and ion emission.  When solids are heated to
temperatures $\gtsimeq \, 500 \, {\rm K}$, they emit electrons and/or
ions.  The emission of electrons depends on the work function of the
solids comprising the grains, which also affects whether evaporating
atoms such as potassium are emitted as neutral atoms or ions.
Including the thermionic and ion emission completely changes the role
of dust grains in the ionization of the gas: instead of mopping up
free charges, they generate them.

In this paper we examine the effects of thermionic and ion emission on
the ionization state of high-temperature, dusty gas.  In \S 2 we
discuss the microphysical mechanisms of thermionic and ion emission,
as well as the evaporation and condensation of K atoms, and provide
quantitative rates and likely input parameters.  In \S 3 we create a
simple model treating evaporation and condensation of K atoms and
ions, including ion emission and thermionic emission.  We show that,
because of the high work function of silicate grains, ion emission
creates signficant ionization at temperatures around 800~K, possibly
as low as 600~K.  In \S 4 we discuss the impact on several aspects of
protoplanetary disks.  In particular, we address where the dead zone's
inner edge should lie, what temperatures are needed for disk gas to
diffuse into the star's magnetosphere, and whether including this new
ionization source can enable the ``short-circuit instability'' of
McNally et al.\ (2013).  Thermionic and ion emission profoundly affect
the ionization state of high-temperature gas in disks.
 
\section{High-Temperature Chemical Network} 

\subsection{High-Temperature Ionization Effects}

In the dusty gas of protoplanetary disks, we cannot always use the
Saha equation, and instead must check the rates of various ionization
and recombination processes.  What's more, protoplanetary disks are
marked by transient temperature changes, and we are interested in how
the ionization responds to the changing conditions.  Thus motivated,
in this section we lay out several processes' kinetic rates and
construct a chemical network to find the steady-state abundances of
electrons and ions as well as the timescales for the ionization
chemistry to reach equilibrium.

Our chemical network includes five species: electrons, molecular or
atomic ions, K atoms bound to dust grains, gas-phase neutral K atoms,
and gas-phase K ions.  Energetic particles (cosmic rays, radioactive
decay products, etc.), and at high temperatures also gas-phase
reactions, ionize the gas molecules; ions and electrons recombine in
the gas phase.  We also include dust grains, calculating their
steady-state charge.  All the other species can collide with and
condense on or chemically react with the grains.  The grains can lose
K atoms at high temperatures, and can eject electrons and ions by the
thermionic effect and by ion emission, respectively.  Before
discussing the chemical network, we review these chemical processes.

\subsection{Gas-Phase Collisional Ionization and Recombination} 

As discussed above, the ${\rm H}_{2}$ molecules' ionization can come
from energetic particles such as Galactic cosmic rays, X-ray photons,
or particles from radioactive decays.  The number of ionizations per
volume per time is defined as $\zeta \, n_{{\rm H}_{2}}$.  We are most
interested in protoplanetary disks' interiors, so we assume that the
cosmic rays and X-rays are shielded, leaving only ionization by the
decay of radioactive elements.  If short-lived radionuclides are not
present, then decay of ${}^{40}{\rm K}$, as well as ${}^{238}{\rm U}$,
${}^{235}{\rm U}$ and ${}^{232}{\rm Th}$, dominate, yielding $\zeta
\approx 6.9 \times 10^{-22} \, {\rm s}^{-1}$ (Umebayashi \& Nakano
1981).  If the short-lived ($t_{1/2} = 0.71 \, {\rm Myr}$)
radionuclide ${}^{26}{\rm Al}$ is present at levels inferred from
meteorites (MacPherson et al.\ 1995), then $\zeta \sim 10^{-18} \,
{\rm s}^{-1}$ (Umebayashi \& Nakano 2009).  Our focus here is on
high-temperature ionization, so we arbitrarily take $\zeta = 6.9
\times 10^{-22} \, {\rm s}^{-1}$ in what follows.  
(If ionizations by ${}^{26}{\rm Al}$ are significant, the ion and 
electron densities would rise sharply above 800 K instead of 700 K.)
The ${\rm H}^+_2$
ions quickly exchange charge with other molecules to produce ions such
as ${\rm HCO}^{+}$, which then also exchange charge with free atoms,
creating ions such as ${\rm Mg}^{+}$.  We assume these steps are
faster than the other reactions, so each ionization of ${\rm H}_{2}$
quickly yields an atomic ion.  We do not consider ${\rm K}^{+}$ as one
of these ions, because other species like Mg are more abundant.
Instead we assume ${\rm K}^{+}$ ions are produced mainly by thermal
ionization.

Atomic ions and free electrons can recombine in the gas phase.  We assume a rate per volume 
$\beta n_{\rm e} n_{\rm i}$, where $\beta = 3 \times 10^{-11} \, T^{-1/2} \, {\rm cm}^{3} \, {\rm s}^{-1}$,
$T$ measured in K (Oppenheimer \& Dalgarno 1974).
Ions and free electrons can also be adsorbed onto grain surfaces, at rates 
\begin{equation}
{\cal R}_{\rm i,ads} = n_{\rm i} \, n_{\rm gr} \, \pi a_{\rm gr}^2 \, ( 8 k T / \pi m_{\rm i})^{1/2} \, S_{\rm i} \, \tilde{J}_{\rm i}
\equiv n_{\rm i} \, \nu_{\rm i,ads},
\end{equation}
and 
\begin{equation} 
{\cal R}_{\rm e,ads} = n_{\rm e} \, n_{\rm gr} \, \pi a_{\rm gr}^2 \, ( 8 k T / \pi m_{\rm e})^{1/2} \, S_{\rm i} \, \tilde{J}_{\rm e}
\equiv n_{\rm e} \, \nu_{\rm e,ads},
\end{equation}
where $\tilde{J}_{\rm i}$ and $\tilde{J}_{\rm e}$ reflect modifications to the collision cross section 
due to the grain charge (Draine \& Sutin 1987). 

Thermal ionization of alkali atoms occurs if they are in the gas phase and suffer collisions with
ambient molecules, mostly ${\rm H}_{2}$.
This process can be represented by the reaction 
\[
{\rm H}_{2} + {\rm K}^{0} \rightarrow {\rm H}_{2} + {\rm K}^{+} + {\rm e}^{-}.
\]
In the gas phase, electrons and ions can radiatively recombine, as follows,
\[
{\rm K}^{+} + {\rm e}^{-} \rightarrow {\rm K}^{0} + \gamma,
\]
but in the regions of protoplanetary disks where temperatures are
$\gtsimeq \, 500 \, {\rm K}$ (the midplane regions well inside 1~AU),
the gas densities are likely high enough that three-body
recombinations dominate:
\[
{\rm H}_{2} + {\rm K}^{+} + {\rm e}^{-} \rightarrow {\rm H}_{2} + {\rm K}^{0}.
\]
We justify this statement below.

Literature on the collisional ionization and recombination rates is
sparse.  Pneumann \& Mitchell (1965) present a kinetic rate, but cite
no measurements or sources.  Their rate also appears to conflict with
that provided by Ashton \& Hayhurst (1973).  Since the latter is
experimentally determined by analyzing the chemical reactions in
flames, it is our preferred value.  Following Ashton \& Hayhurst
(1973), collisional ionization takes place at a rate per unit volume
\begin{equation}
{\cal R}_{\rm gas,ion} = k_2 \, n_{{\rm H}_{2}} \, n_{ {\rm K}^{0} },
\end{equation}
where 
\begin{equation} 
k_2 = (9.9 \pm 2.7) \times 10^{-9} \, T^{1/2} \, \exp \left(
-\frac{\rm IP}{k T} \right) \, {\rm cm}^{3} \, {\rm s}^{-1},
\end{equation}
and $T$ is measured in K.  Here ${\rm IP} = 4.34 \, {\rm eV}$ for K,
and ${\rm IP} / k = 50350 \, {\rm K}$.  The 3-body gas-phase
recombination rate is
\begin{equation}
{\cal R}_{\rm gas, rec} = k_{-2} \, n_{{\rm H}_{2}} \, n_{ {\rm K}^{+} } \, n_{\rm e} 
\end{equation}
where 
\begin{equation}
k_{-2} = (4.4 \pm 1.1) \times 10^{-24} \, T^{-1} \, {\rm cm}^{6} \, {\rm s}^{-1},
\end{equation}
and again $T$ is measured in Kelvin (Ashton \& Hayhurst 1973). 

The two coefficients $k_2$ and $k_{-2}$ must be related to each other.
In the absence of dust, these would be the only rates affecting the
ionization fraction, and should be in detailed balance with each other
in local thermodynamic equilibrium.  The ionization fractions found
using these coefficients thus should conform to the Saha equation.
Hence, ${\cal R}_{\rm gas, rec} = {\cal R}_{\rm gas, ion}$, and $n_{
  {\rm K}^{+} } \, n_{\rm e} \, k_{-2} = n_{ {\rm K}^{0} } \, k_2$,
and therefore we expect
\begin{equation}
\frac{ k_{2} }{ k_{-2} } =
\frac{ n_{{\rm K}^{+}} n_{\rm e} }{ n_{{\rm K}_{0}} } = 
\frac{ 2 g_{+} }{ g_0 } \, \left( \frac{ 2\pi m_{\rm e} k T }{ h^2 } \right)^{3/2} \, \exp \left( -\frac{\rm IP}{k T} \right) 
 = 2.415 \times 10^{15} \, T^{3/2} \, \exp \left( -\frac{\rm IP}{k T} \right) 
\end{equation}
The values from Ashton \& Hayhurst (1973) yield 
\begin{equation}
\frac{ k_{2} }{ k_{-2} } =
  (2.3 \pm 0.8) \times 10^{15} \, T^{3/2} \, \exp \left( -\frac{\rm IP}{k T} \right),
\end{equation}
consistent with the Saha equation.

Recombinations by 3-body reactions proceed at a rate $(k_{-2} \,
n_{{\rm H}_{2}}) \, n_{{\rm K}^{+}} \, n_{\rm e}$, which is to be
compared to the rate of radiative 2-body recombinations, $\beta \,
n_{{\rm K}^{+}} \, n_{\rm e}$.  For typical radiative recombinations,
$\beta \sim 3 \times 10^{-11} \, T^{-1/2} \, {\rm cm}^{3} \, {\rm
  s}^{-1}$ (Oppenheimer \& Dalgarno 1974).  At $T = 500 \, {\rm K}$,
$k_{-2} \approx 10^{-26} \, {\rm cm}^{6} \, {\rm s}^{-1}$, and if
$n_{{\rm H}_{2}} = 10^{14} \, {\rm cm}^{-3}$, then $k_{-2} \, n_{{\rm
    H}_{2}} \sim 10^{-12} \, {\rm cm}^{3} \, {\rm s}^{-1}$.  For
densities $n_{{\rm H}_{2}} \, \gtsimeq \, 10^{14} \, {\rm cm}^{-3}$,
radiative recombinations of ${\rm K}^{+}$ and electrons are less significant
than 3-body recombinations, but we include both rates in our
calculations. 

\subsection{Thermionic Emission} 

Thermionic emission, the ejection of electrons from heated solids, is
a long recognized effect and the basis for the cathode ray tube.  The
rate at which electrons are emitted, per surface area of solid, is
described by Richardson's law,
\begin{equation}
j(T) = \lambda_{\rm R} \, \frac{ 4\pi m_{\rm e} (k T)^2 }{ h^3 } \, \exp \left( -\frac{W}{k T} \right),
\end{equation}
where $\lambda_{\rm R}$ is a dimensionless number of order unity
depending on the solid material, and is typically $\approx 1/2$
(Crowell et al. 1965).

Thermionic emission is sensitive to $W$, the work function of the
material -- that is, the energy needed to make an electron escape the
solid.  As the grains lose electrons and develop a positive charge $+Z
e$, removing more electrons requires extra work to overcome the
Coulomb force.  The effective work function is
\begin{equation}
W_{\rm eff} = W + \frac{ Z e^2 }{a_{\rm gr} }.
\end{equation}
For conductors, the escape takes place from the top of the conduction
band, and the work function is easily measured and commonly tabulated.
For example, the work function of graphite is 4.62~eV (Jain \&
Krishnan 1952), Fe has a work function of 4.31~eV, and that of Ni is
4.50~eV (Fomenko 1966).

Unfortunately, the grains likely to occur in the solar nebula and
other protoplanetary disks contain many insulating materials, making
determining the work function difficult.  The best analogs to the
grains in the solar nebula, and presumably in other protoplanetary
disks as well, are the unaltered (pre-accretionary) matrix grains in
chondrites.  These grains are composed of crystalline, Mg-rich
olivines [${\rm Mg}_{2}{\rm SiO}_{4}$] and pyroxenes [${\rm
    MgSiO}_{3}$]; relatively Fe-rich amorphous silicates [$({\rm
    Mg,Fe})_{2}{\rm SiO}_{4}$] and [$({\rm Mg,Fe}){\rm SiO}_{3}$]; and
a smaller fraction (a few percent) of grains of metallic Fe,Ni and
their sulfides (Scott \& Krot 2005).  Metallic FeNi grains are
conducting, with a work function presumably $\approx 4.4$~eV, but the
silicates are insulators.  Because of this, work functions of
silicates are difficult to measure, and few data exist in the
literature.  Quartz (${\rm SiO}_2$) has a work function 5.0~eV, and
mica (a phyllosilicate) has a work function 4.8~eV (Fomenko 1966), but
the authors were unable to find any other direct measurements for
specific silicate minerals.  Far better than quartz and mica as
analogs for solar nebula dust, however, is the JSC-1 lunar regolith
simulant (McKay et al.\ 1993), composed of a mixture of 40 wt\%
ferromagnesian olivine [$({\rm Fe,Mg})_{2}{\rm SiO}_{4}$], 40 wt\%
plagioclase [${\rm NaAlSi}_{3}{\rm O}_{8}$ and ${\rm CaAl}_{2}{\rm
    Si}_{2}{\rm O}_{8}$], the remainder comprised of Fe- and Ti-rich
minerals like ilmenite (${\rm FeTiO}_3$), anatase [${\rm TiO}_{2}$],
magnetite [${\rm Fe}_{3}{\rm O}_{4}$], hematite [${\rm Fe}_{2}{\rm
    O}_{3}$] and pseudobrookite [${\rm Fe}_{2}{\rm TiO}_{5}$].
Trigwell et al.\ (2009) examined how JSC-1 simulant is charged when
brought into contact with different substances, and found it was
negatively charged by Al (work function 4.28~eV), Cu (4.65~eV) and
stainless steel (5.04~eV), but positively charged by the organic
polymer PTFE (work function 5.75~eV).  The acquired charge correlates
with the work function of the other material, and the correlation
strongly suggests JSC-1 itself has an effective work function
somewhere between 5.0 and 5.4~eV.  Feuerbacher et al.\ (1972) examined
the photoelectric effect using JSC-1 simulant and found the threshold
for photoelectron emission implies an effective work function close to
5.0~eV.  Based on these measurements, we adopt a work function of
5.0~eV for solar nebula dust grains.

% NB: Note that this disagrees with an implied work function based on
% placement in the triboelectric series.  Quartz (${\rm SiO}_2$) and
% mica (a phyllosilicate) are minerals that generally become
% positively charged upon contact electrification with other
% materials, even Ca, with a work function 2.8~eV.  This implies that
% they have work functions $\approx 2.5$~eV or less; but this
% conflicts with actual measurements.  Also note that Weingartner \&
% Draine 2001 inferred $W = 8$~eV from the Feuerbacher et al.\ (1972)
% data, but the source paper makes clear they prefer a work function
% of 4.97~eV.

With the work function defined, the rate of electron production per
volume due to thermionic emission from dust is
\begin{equation}
{\cal R}_{\rm therm} = n_{\rm gr} \, 4\pi a_{\rm gr}^2 \, j(T) = 
                n_{\rm gr} \, 4\pi a_{\rm gr}^2 \, \lambda_{\rm R} \, 
                \frac{ 4\pi m_{\rm e} (k T)^2 }{ h^3 } \, \exp \left( -\frac{W_{\rm eff}}{k T} \right).
\end{equation} 
Again, for silicates we adopt $W = 5.0 \, {\rm~eV}$.  Since data are
lacking for the coefficient $\lambda_{\rm R}$, we adopt $\lambda_{\rm
  R} = 1/2$.

The inverse of thermionic emission is electrons sticking on dust
grains, which occurs at a rate $n_{\rm e} \, \nu_{\rm ads}$.  In the
absence of gas-phase processes affecting electrons, thermionic
emission and its inverse could be expected to be in detailed balance,
with ${\cal R}_{\rm therm} = {\cal R}_{\rm e,ads}$ and
\begin{equation}
\lambda_{\rm R} \, \frac{4\pi m_{\rm e} k^2 T^2}{h^3} \, \exp \left( -\frac{W_{\rm eff}}{k T} \right) =
 \tilde{J}_{\rm e} \, S_{\rm e} \, n_{\rm e} \, \frac{1}{4} \left( \frac{8 k T}{\pi m_{\rm e}} \right)^{1/2} 
\end{equation}
Note that since the electrons' production and destruction rates both
scale with the grain surface area, the grain density and radii drop out of the balance
equation.  Solving for $n_{\rm e}$ yields
\begin{equation}
n_{\rm e} =
 \frac{ 4 \lambda_{\rm R} }{ 2 \tilde{J}_{\rm e} \, S_{\rm e} } \, 
 \left( \frac{ 2\pi m_{\rm e} k T }{ h^2 } \right)^{3/2} \, \exp \left( -\frac{W_{\rm eff}}{k T} \right).
\end{equation} 
This is identical in form to a Saha equation, involving a charged and
neutral species $X^{+}$ and $X^{0}$ with statistical weights $g^{+}$
and $g^{0}$, provided $(n_{{\rm X}^{0}} / n_{{\rm X}^{+}}) (g_{+} /
g_{0}) = \lambda_{\rm R} / (2 \tilde{J}_{\rm e} \, S_{\rm e}) \sim 1$.
Potassium ion and electron densities essentially follow a Saha
equation, but with an energy equal to $W_{\rm eff}$, which in practice
is typically in the range of 3 to 4~eV (see below), {\it not} the
ionization potential ${\rm IP} = 4.34 \, {\rm eV}$.  In this way the
dust grains fundamentally alter the gas ionization state, even in the
high-temperature limit, where their effects are described by a
Saha-like equation.

\subsection{Evaporation of Neutral and Ionized Potassium Atoms}

In addition to electrons, atoms can also emit ions.  For the reasons
outlined above, most relevant are ${\rm K}^{+}$ ions.  At high
temperatures, potassium exists mostly in the gas phase, with K atoms
that meet dust grains quickly vaporizing because of thermal
vibrations.  A single K atom bound to a grain surface vibrates on the
lattice with a frequency $\nu$.  We adopt a typical value $\nu = 3.7
\times 10^{13} \, {\rm s}^{-1}$, appropriate for K bound to Pt
(Hagstrom et al.\ 2000).  In any vibration cycle, there is a
probability of vaporization $\exp(-E_{\rm a} / kT)$, where $E_{\rm a}$
is the activation energy, equal to the binding energy, typically
several eV.  Defining $n_{\rm K,cond}$ as the number of K atoms in the
grain minerals per volume of nebula, the vaporization rate per volume
is
\begin{equation}
{\cal R}_{\rm K,evap} = n_{\rm K,cond} \, \nu \, \exp \left( -\frac{ E_{\rm a} }{ k T } \right) 
 \equiv n_{\rm K,cond} \, \nu_{\rm evap}.
\end{equation} 
Balancing these vaporizations is the condensation of K atoms onto the
grain surfaces.  The total rate at which K atoms collide with dust
grains is ${\cal R}_{\rm K0,coll} + {\cal R}_{\rm K+,coll}$, where
\begin{equation}
{\cal R}_{\rm K+,coll} = n_{{\rm K}^{+}} \, n_{\rm gr} \, \pi \, a_{\rm gr}^{2} \, \left( \frac{8 k T}{\pi m_{\rm K}} \right)^{1/2} \,
 \tilde{J}_{\rm i} \equiv n_{{\rm K}^{+}} \, \nu_{{\rm K}^{+},{\rm coll}},
\end{equation} 
$\tilde{J}_{\rm i}$ being the focusing factor appropriate for ions
(Draine \& Sutin 1987) and $m_{\rm K}$ the mass of a potassium atom,
and
\begin{equation}
{\cal R}_{\rm K0,coll} = n_{{\rm K}^{0}} \, n_{\rm gr} \, \pi \, a_{\rm gr}^{2} \, \left( \frac{8 k T}{\pi m_{\rm K}} \right)^{1/2} 
\equiv n_{{\rm K}^{0}} \, \nu_{{\rm K}^{0},{\rm coll}}.
\end{equation} 
The ratio of bound K atoms to gas-phase K atoms can be found be
balancing these two rates:
\[
\frac{ n_{\rm K,cond} }{ n_{\rm K} } \sim n_{\rm gr} \, \pi a_{\rm gr}^{2} \, 
                                         \left( \frac{8 k T}{\pi m_{\rm K}} \right)^{1/2} \, S_{\rm i} \, \nu^{-1} \, 
                                         \, \exp \left( +\frac{ E_{\rm a} }{ k T } \right)  
\]
\begin{equation}
 = \rho_{\rm g} \, \frac{ (\rho_{\rm gr} / \rho_{\rm g}) }{ 4 \rho_{\rm s} a_{\rm gr} / 3 } \, 
                                         \left( \frac{8 k T}{\pi m_{\rm K}} \right)^{1/2} \, S_{\rm i} \, \nu^{-1} \, 
                                         \, \exp \left( +\frac{ E_{\rm a} }{ k T } \right).
\end{equation}
The fraction of potassium atoms in the gas phase (versus bound in
grains) is strongly temperature dependent.

At low temperatures, potassium is unlikely to be found in the gas
phase because it will chemically react with silicates and be bound
into minerals.  In particular, potassium resides in feldspar (${\rm
  KAlSi}_{3}{\rm O}_{8}$) below its 50\% condensation temperature
$\approx 1006 \, {\rm K}$ (Lodders 2003).  To mimic this behavior, we
assume the binding energy of K atoms in silicates is $E_{\rm a} = 3.26
\, {\rm eV}$, so that $n_{\rm K,coll} / n_{\rm K} = 1/2$ at $T = 1006
\, {\rm K}$.  Here we take a gas density $n_{{\rm H}_{2}} = 10^{14} \,
   {\rm cm}^{-3}$, a dust-to-gas ratio $= 0.01$, a grain density
   $\rho_{\rm s} = 3 \, {\rm g} \, {\rm cm}^{-3}$, a grain radius
   $a_{\rm gr} = 1 \, \mu{\rm m}$, a sticking coefficient $S_{\rm i} =
   1$, and neglecting the charge state of the K atoms.

At temperatures above 1006~K, potassium atoms will evaporate from the
surface, but whether they leave as neutral atoms or as ions depends on
their ionization potential and the work function of the grains,
according to the Saha-Langmuir equation.  Among departing potassium
atoms, the ratio of ions to neutrals obeys
\begin{equation} 
\frac{n_{ {\rm K}^{+} }}{n_{ {\rm K}^{0} }} = \frac{ g_{+} }{ g_{0} } \, 
\exp \left( +\frac{ {W_{\rm eff}} - {\rm IP} }{ k T } \right).
\end{equation} 
The fraction of leaving K atoms that are positive ions is then 
\begin{equation}
f_{+} = \frac{ ( n_{{\rm K}^{+}} / n_{{\rm K}^{0}} ) }
             { 1 + ( n_{{\rm K}^{+}} / n_{{\rm K}^{0}} ) }.
\end{equation}
The rate at which ions evaporate from grains (per volume of nebula) is
therefore ${\cal R}_{\rm K,evap} f_{+}$, and the rate at which neutral
K atoms do so is ${\cal R}_{\rm K,evap} (1 - f_{+})$.

\subsection{Chemical Network}

We now construct a chemical network to calculate the densities of the
five species, $n_{\rm e}$ (free electrons), $n_{\rm i}$ (gas-phase
atomic ions), $n_{\rm K,coll}$ (grain-adsorbed K atoms), $n_{{\rm
    K}^{+}}$ (gas-phase K ions), and $n_{{\rm K}^{0}}$ (gas-phase
neutral K atoms), as well as the charge $Z$ on each dust grain.
% We assume a monodispersion of grain sizes, $a_{\rm gr} = 1 \,
% \mu{\rm m}$, a dust-to-gas ratio 0.01, an internal grain density
% $\rho_{\rm s} = 3 \, {\rm g} \, {\rm cm}^{-3}$,
The time derivatives of these quantities are as follows.
\begin{eqnarray}
\frac{d n_{\rm i}}{dt}        & = & +\zeta \, n_{{\rm H}_{2}} -\beta n_{\rm e} n_{\rm i} -{\cal R}_{\rm i,ads} \\
\frac{d n_{\rm e}}{dt}        & = & +\zeta \, n_{{\rm H}_{2}} -\beta n_{\rm e} n_{\rm i} -{\cal R}_{\rm e,ads} 
                                    -{\cal R}_{\rm gas,rec} +{\cal R}_{\rm gas,ion} +{\cal R}_{\rm therm} 
 \label{eq:ne} \\ 
\frac{d n_{{\rm K}^{+}} }{dt} & = & -{\cal R}_{\rm gas,rec} +{\cal R}_{\rm gas,ion} +{\cal R}_{\rm K,evap} \, f_{+} \,  
                                    -{\cal R}_{\rm K+,coll} \\ 
\frac{d n_{{\rm K}^{0}} }{dt} & = & +{\cal R}_{\rm gas,rec} -{\cal R}_{\rm gas,ion} 
                                        +{\cal R}_{\rm K,evap} \, (1 - f_{+}) \,  -{\cal R}_{\rm K0,coll} \\ 
\frac{d n_{{\rm K,cond}} }{dt} & = & +{\cal R}_{\rm K+,coll} +{\cal R}_{\rm K0,coll} -{\cal R}_{\rm K,evap}
\end{eqnarray} 
It is straightforward to confirm that the total number of potassium
atoms (neutral, ionized, and bound) is conserved in such a model,
i.e., $n_{\rm K,cond} + n_{{\rm K}^{+}} + n_{{\rm K}^{0}} = n_{\rm
  K,tot}$, where we assume $n_{\rm K,tot} = 3.04 \times 10^{-7} \,
n_{{\rm H}_{2}}$ (Lodders 2003).  We seek a steady-state solution by
setting all time derivatives to zero.  Note that the five densities
are, through the cross section modifications $\tilde{J}$, all
functions of the grain charge $Z$, the sixth variable.  The sixth
equation needed to close the system corresponds to charge neutrality.
If grains' mean charge is $+Ze$,
\begin{equation}
n_{\rm gr} Z + n_{{\rm K}^{+}} +n_{{\rm Mg}^{+}} - n_{\rm e} = 0.
\end{equation}
In practice, we vary $Z$ till charge neutrality is achieved and the
other densities are in steady state.

More specifically, we find the steady-state solution for gas of
density $n_{{\rm H}_{2}}$ and temperature $T$ as follows.  At a
particular value of $Z$, we calculate the cross section modifications
for ions, electrons, and neutral atoms, using the approximation
formulas of Draine \& Sutin (1987).  
We assume the gas and dust temperatures are identical; in protoplanetary disks this is an 
excellent assumption except for the uppermost $\ltsimeq \, 0.1 \, {\rm g} \, {\rm cm}^{-2}$
surface layers (Glassgold et al.\ 2004).
With the coefficients known, we
then solve for the species' steady-state abundances.  Given guesses
for the densities of ${\rm K}^{0}$ and ${\rm K}^{+}$, we find the
electron density by first solving for $n_{\rm i}$:
\begin{equation}
n_{\rm i} = \frac{ \zeta \, n_{{\rm H}_{2}} }{ \beta \, n_{\rm e} + n_{\rm gr} \, \nu_{\rm, i,ads} }
          = \frac{ \zeta \, n_{{\rm H}_{2}} }{ \beta \, n_{\rm e} + n_{\rm gr} \, \pi a_{\rm gr}^{2} \, 
                C_{\rm i} \, S_{\rm i} \, \tilde{J}_{\rm i} }.
\end{equation}
This value is then substituted into the equation for the electron density, 
\begin{equation}
%\zeta n_{{\rm H}_{2}} + {\cal R}_{\rm therm} + k_{2} \, n_{{\rm H}_{2}} \, n_{{\rm K}^{0}} = 
%  n_{\rm e} \, \left[ \beta \, n_{\rm i} +k_{-2} \, n_{{\rm H}_{2}} \, n_{{\rm K}^{+}}
% +n_{\rm gr} \, \pi a_{\rm gr}^2 \, C_{\rm e} \, S_{\rm e} \, \tilde{J}_{\rm e} \right], 
\zeta n_{{\rm H}_{2}} + {\cal R}_{\rm therm} + k_{2} \, n_{{\rm H}_{2}} \, n_{{\rm K}^{0}} = 
  n_{\rm e} \, \left[ \beta \, n_{\rm i} +k_{-2} \, n_{{\rm H}_{2}} \, n_{{\rm K}^{+}} +\nu_{\rm e,ads} \right], 
\end{equation}
to derive a quadratic equation for $n_{\rm e}$:
\[
% \left[ \beta \, \pi a_{\rm gr}^{2} \, C_{\rm e} \, S_{\rm e} \, \tilde{J}_{\rm e} 
%       +\beta \, k_{-2} \, n_{{\rm H}_{2}} \, n_{{\rm K}^{+}} \right] \, n_{\rm e}^{2} 
 \left[ \beta \, \left( \nu_{\rm e,ads} +k_{-2} \, n_{{\rm H}_{2}} \, n_{{\rm K}^{+}} \right) \right] \, n_{\rm e}^{2} 
\]
\[
%+\left[ n_{\rm gr} \, \pi a_{\rm gr}^{2} \, C_{\rm i} \, S_{\rm i} \tilde{J}_{\rm i} \, 
%            \left( \pi a_{\rm gr}^{2} \, C_{\rm e} \, S_{\rm e} \tilde{J}_{\rm e} 
%                     +k_{-2} \, n_{{\rm H}_{2}} \, n_{{\rm K}^{+}} \right) 
%       -\beta \, \left( 4\pi a_{\rm gr}^{2} \, j(T) + k_{2} \, n_{{\rm H}_{2}} \, n_{{\rm K}^{0}} \right) 
%  \right] \, n_{\rm e} 
+\left[ \nu_{\rm i,ads} \, \left( \nu_{\rm e,ads} + k_{-2} \, n_{{\rm H}_{2}} \, n_{{\rm K}^{+}} \right)
 -\beta \, \left( n_{\rm gr} \, 4\pi a_{\rm gr}^{2} \, j(T) + k_{2} \, n_{{\rm H}_{2}} \, n_{{\rm K}^{0}} \right) \right] \, n_{\rm e}
\]
\begin{equation}
%-\left( \pi a_{\rm gr}^{2} \, C_{\rm i} \, S_{\rm i} \, \tilde{J}_{\rm i} \right) \, 
% \left[ \zeta \, n_{{\rm H}_{2}} + n_{\rm gr} \, 4\pi a_{\rm gr}^{2} \, j(T) 
%            +k_{2} \, n_{{\rm H}_{2}} \, n_{{\rm K}^{0}} \right] = 0.
-\nu_{\rm i,ads} \, \left[ \zeta \, n_{{\rm H}_{2}} + n_{\rm gr} \, 4\pi a_{\rm gr}^{2} \, j(T) 
 +k_{2} \, n_{{\rm H}_{2}} \, n_{{\rm K}^{0}} \right] = 0.
\end{equation}
Once $n_{\rm e}$ is found, $n_{\rm i}$ is found using the above
equation, as are the other densities, as follows:
\begin{equation}
n_{{\rm K}^{+}} =
 \frac{ n_{\rm K,tot} \, \left( k_{2} \, n_{{\rm H}_{2}} +\nu_{{\rm K}^{0},{\rm coll}} \, f^{+} \right) }
{ k_{2} \, n_{{\rm H}_{2}} \, \left( 1 + \frac{\nu_{{\rm K}^{+},{\rm coll}}}{\nu_{\rm evap}} \right)
 +k_{-2} \, n_{{\rm H}_{2}} \, n_{\rm e} \, \left( 1 + \frac{\nu_{{\rm K}^{0},{\rm coll}}}{\nu_{\rm evap}} \right) 
 +\nu_{{\rm K}^{+},{\rm coll}} \, \left( 1 - f^{+} \right) 
 +\nu_{{\rm K}^{0},{\rm coll}} \, f^{+} 
 +\frac{ \nu_{{\rm K}^{0},{\rm coll}} \, \nu_{{\rm K}^{+},{\rm coll}} }{ \nu_{\rm evap} }  }
\end{equation} 
yields the ${\rm K}^{+}$ density, in terms of which 
\begin{equation} 
n_{{\rm K}^{0}} = 
\frac{ n_{\rm K,tot} - n_{{\rm K}^{+}} \, \left( 1 + \frac{ \nu_{{\rm K}^{+},{\rm coll}} }{ \nu_{\rm evap} } \right) }
     {                                           1 + \frac{ \nu_{{\rm K}^{0},{\rm coll}} }{ \nu_{\rm evap} }         }, 
\end{equation} 
and
\begin{equation} 
n_{\rm K,cond} = 
\frac{ n_{\rm K,tot} - n_{{\rm K}^{+}} \, \left( 1 - \frac{ \nu_{{\rm K}^{+},{\rm coll}} }{ \nu_{{\rm K}^{0},{\rm coll}} } \right) }
     {                                           1 + \frac{ \nu_{\rm evap} }{ \nu_{{\rm K}^{0},{\rm coll}} }                       }.
\end{equation} 
In practice, at each guess for $Z$ we iterate between solving for
$n_{\rm e}$ and solving for $n_{\rm i}$, $n_{{\rm K}^{0}}$ and
$n_{{\rm K}^{+}}$, then updating $n_{\rm e}$.  Performing 10 such
iterations allows us to compute the net charge $n_{\rm i} + n_{{\rm
    K}^{+}} +n_{\rm gr} Z - n_{\rm e}$ for the given $Z$.  At very
large $Z > 0$ the net charge is positive, while at very large $Z < 0$
it is negative, so it goes to zero at some finite $Z$.  We use a
bisection method with 80 steps to find the value of $Z$ that yields
both charge neutrality and ionization equilibrium.

\section{Results} 

\subsection{Canonical Case}

We have solved for the abundances of electrons, ions (${\rm Mg}^{+}$
and ${\rm K}^{+}$), as well as neutral and bound K atoms, using the
chemical network described above.  We start with our canonical case,
characterized by $n_{{\rm H}_{2}} = 10^{14} \, {\rm cm}^{-3}$, grains
with a dust-to-gas mass ratio 0.01, uniform radius $a_{\rm gr} = 1 \,
\mu{\rm m}$, and work function $W = 5.0 \, {\rm eV}$.
%Figure~\ref{fig:xeCanonical} 
Figure~2 shows the fractional abundances of electrons, $x_{\rm e} =
n_{\rm e} / n_{{\rm H}_{2}}$, as well as $x_{\rm i}$, $x_{{\rm K}^{+}}$, 
$x_{{\rm K}^{0}}$, and $x_{{\rm K,cond}}$, as functions
of temperature.  In cold gas, $\ltsimeq \, 500 \, {\rm K}$, the ion
and electron densities come from the balance between ionizations by
energetic particles and collisions with dust grains.  Because of their
slower thermal velocities, ions take longer to reach the grains and
are more abundant, with $x_{\rm i} \sim 10^{-19}$, compared to $x_{\rm
  e} \sim 10^{-21}$ for electrons.  The very low abundances are a
consequence of the low ionization rates associated with the
radioactivities.  Essentially all potassium is bound into solids, with
few ions or neutral atoms in the gas phase.  As the temperature
increases, however, ${\rm K}^{0}$ and especially ${\rm K}^{+}$ enter
the gas phase.  Some come directly from evaporation of ${\rm K}^{+}$
ions from grain surfaces; because ${\rm IP} < {W}$, most K atoms leave
as ions.  In addition, once the temperatures exceed about 750~K,
electrons are produced effectively by thermionic emission.  Above
1000~K, the potassium atoms evaporate from solids; most takes the form
of neutral ${\rm K}^{0}$ atoms in the gas-phase.  Only in gas hotter
than about 1500~K do the ${\rm K}^{+}$ ions compare in abundance to
the neutral ${\rm K}^{0}$ atoms.
%-------------------------------------------------------------------------------
%
% FIGURE 2 
%
\begin{figure}[H]
\begin{center} 
\includegraphics[width=0.80\paperwidth]{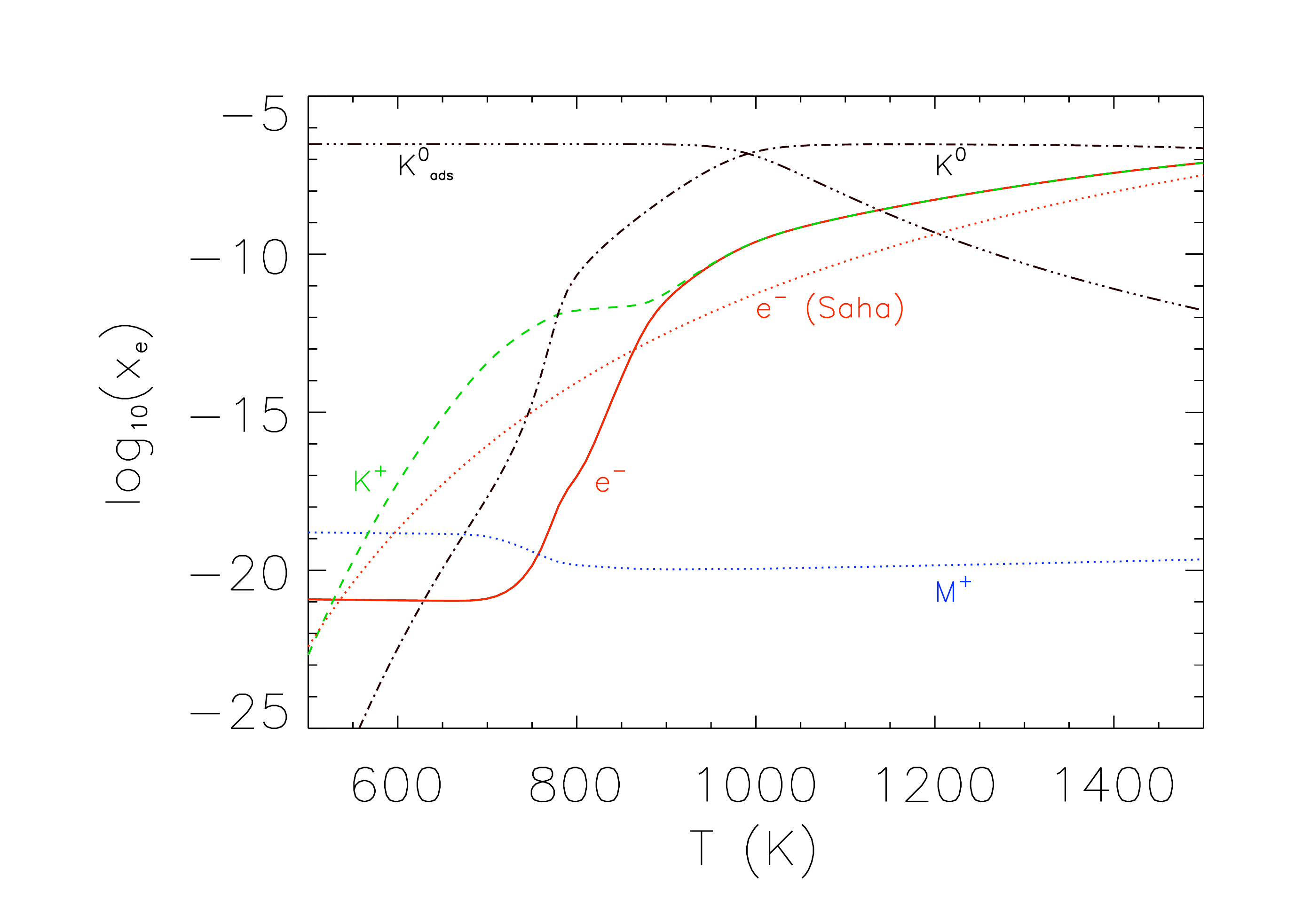}
\caption{Abundances of electrons (solid red curve), 
 atomic (Mg) ions (dotted blue curve), 
 ${\rm K}^{+}$ ions (dashed green curve), 
 gas-phase ${\rm K}^{0}$ atoms (dash-dot curve), and bound K atoms
 (dash-dot-dot-dot curve), as functions of temperature $T$, for the
  canonical case (gas density $n_{{\rm H}_{2}} = 10^{14} \, {\rm cm}^{-3}$, 
 dust-to-gas mass ratio 0.01, grain work function $W = 5.0 \, {\rm eV}$.  
 At low temperatures, non-thermal ionizations
  (from radioisotopes) dominate, but emission of ${\rm K}^{+}$ ions
  from dust grains is significant for $T \, \gtsimeq \, 550 \, {\rm K}$.  
 Evaporation of ${\rm K}^{0}$ atoms from grains, and their
  subsequent thermal ionization, contribute significantly to ${\rm K}^{+}$ 
  and electron densities for $T \, \gtsimeq \, 750 \, {\rm K}$.  
  For $T \, \gtsimeq \, 1000 \, {\rm K}$, thermal ionizations
  and recombinations dominate and are nearly in thermodynamic
  equilibrium.  For comparison, we also plot the electron 
  abundance derived using the Saha equation (dashed red curve). The Saha equation 
  overpredicts the ionization at temperatures $\ltsimeq 800$ K, and underpredicts
  the ionization at higher temperatures.  } 
\end{center}
\label{fig:xeCanonical}
\end{figure}
%
%-------------------------------------------------------------------------------

The bolded curve in %Figure~\ref{fig:ZCanonical}
Figure~3 shows the charge on dust grains as a function of temperature.
Below about 700~K, when the ionization rate is low, grains typically
hold only a few electron charges.  The value of $Z$ is such that
charging by electron and ion collisions balance each other, including
the effects of charge on the collision rates (i.e., $\tilde{J}_{\rm
  e}$, $\tilde{J}_{\rm i}$).  As the temperature increases, however,
ion emission comes to dominate the charging, and the grains quickly
become negatively charged; at 900~K the charge is about $-700 \, e$.
Because adsorption of gas-phase ions is unimportant, the grains charge
until $W_{\rm eff}$ is reduced enough that ion emission is suppressed
and thermionic emission enhanced.  In other words, ions are emitted
until $\left| Z e^2 / a_{\rm gr} {W} \right|$ approaches unity, which
requires $Z \sim 3500$.  At 900~K, $\left| Z e^2 / a_{\rm gr} \right|
\approx 0.2 \, {W} $.  As the temperature increases, collisions of
ions with grain surfaces begin to matter, and the grain charge is
reduced, approaching the thermodynamic limit where $\left| Z e^2 /
a_{\rm gr} \right| \sim k T$, or $Z \sim -100$ (at 1500~K).
%-------------------------------------------------------------------------------
%
% FIGURE 3 
%
\begin{figure}[H]
\begin{center} 
\includegraphics[width=0.80\paperwidth]{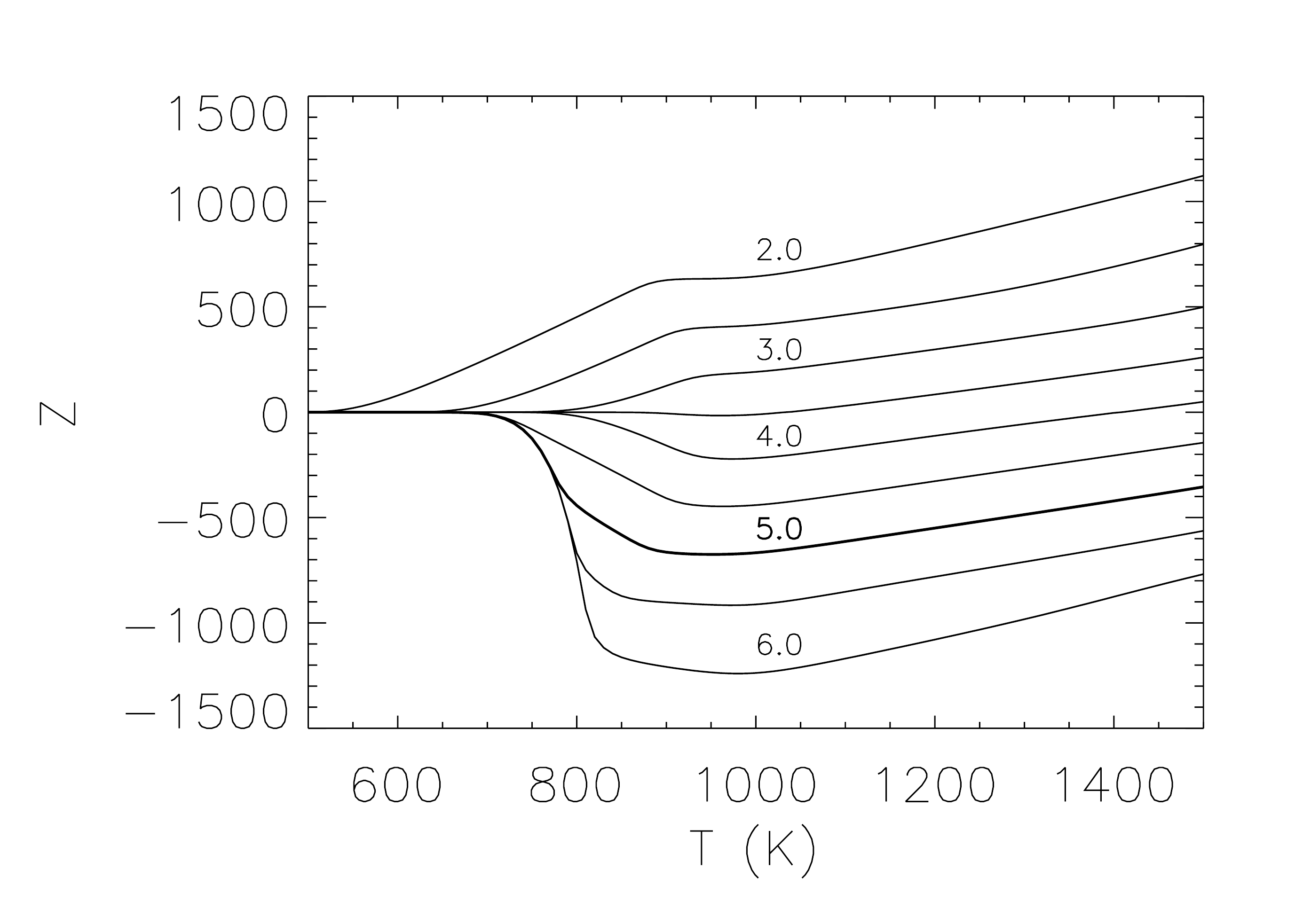}
\caption{Charge on dust grains (in units of elementary charge $e$) as
  a function of temperature, for the canonical case but with various
  grain work functions.  From top to bottom, $W = 2.0$, $2.5$,
  ... $6.0 \, {\rm eV}$.  The bolded curve corresponds to $W = 5.0 \,
  {\rm eV}$.  For low work functions ($W \, \ltsimeq \, 4 \, {\rm
    eV}$, thermionic emission dominates the charging, and dust grains
  are positively charged; as temperature increases, thermionic
  emission becomes more effective, and grains become more positively
  charged.  For large work functions ($W \, \gtsimeq \, 4 \, {\rm
    eV}$), ion emission dominates, and grains become negatively
  charged; as the temperature increases, collisions of ${\rm K}^{+}$
  ions onto the grains becomes more effective, and grains become less
  negatively charged. }
\end{center}
\label{fig:ZCanonical}
\end{figure}
%
%-------------------------------------------------------------------------------

The relative abundances of neutral and ionized K atoms also approach
the thermodynamic limit only at high temperatures.  The bolded curve
in %Figure~\ref{fig:ECanonical}
Figure~4 shows what effective ionization potential energy $E$ must
exist so that the Saha equation would yield the correct ratio for
$n_{{\rm K}^{+}} n_{\rm e} / n_{{\rm K}^{0}}$, i.e.,
\begin{equation}
E \equiv -\left[ \ln \left( n_{{\rm K}^{+}} \, n_{\rm e} / n_{{\rm K}^{0}} \right) -35.42 \right] \, k T.
\end{equation}
At high temperatures it tends to approach the actual value of ${\rm
  IP}$ for potassium, $4.34 \, {\rm eV}$, but at lower temperatures
$E$ is lower, reflecting the fact that many K atoms evaporate as ${\rm
  K}^{+}$ ions from grain surfaces, according to the Saha-Langmuir
equation.  At all temperatures, but especially below 800~K, emissions
from dust grains generate more ${\rm K}^{+}$ ions than the Saha
equation would indicate.
%-------------------------------------------------------------------------------
%
% FIGURE 4 
%
\begin{figure}[H]
\begin{center} 
\includegraphics[width=0.80\paperwidth]{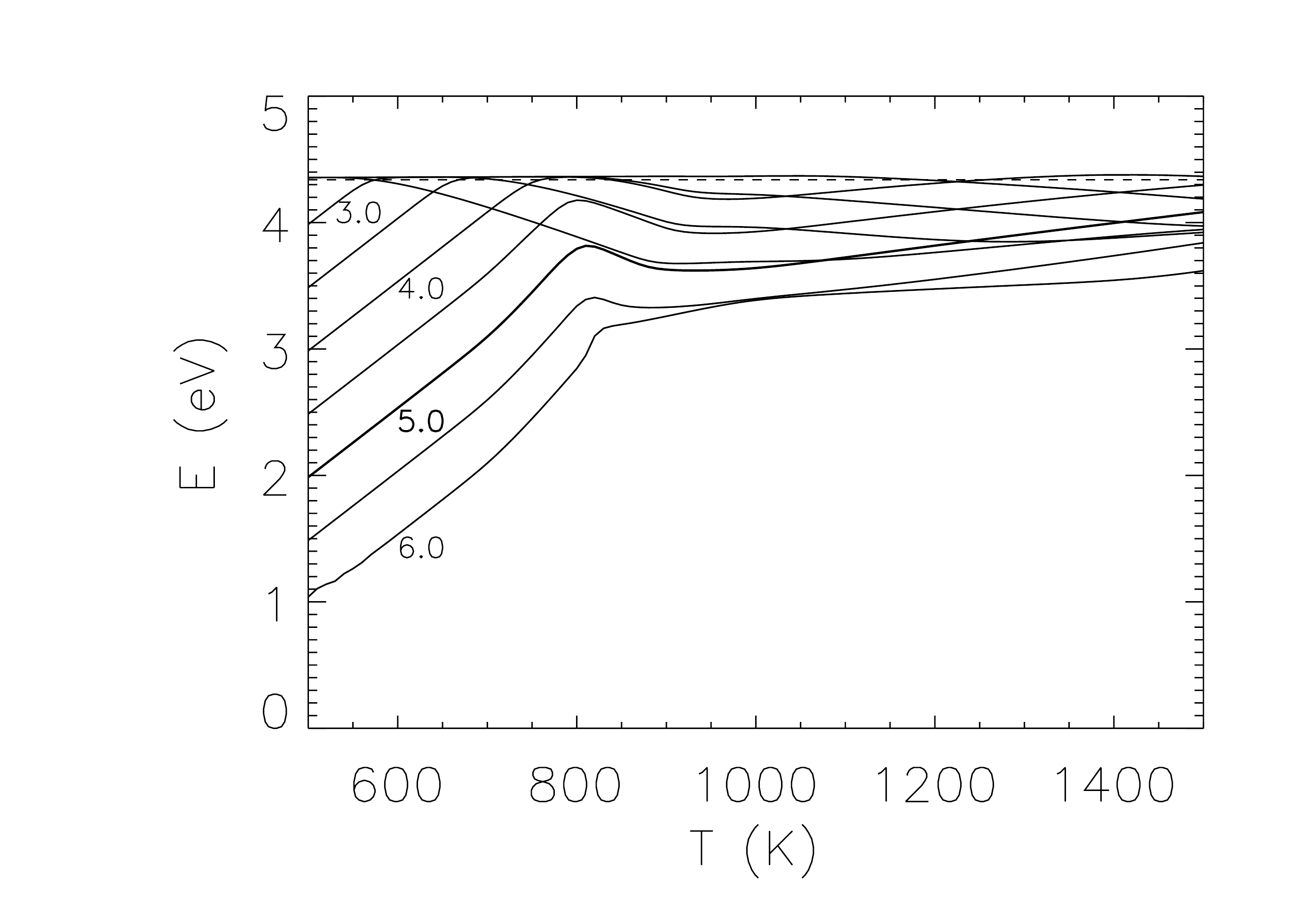}
\caption{Effective ionization potential, $E$, needed for the Saha
  equation to return the correct value of $n_{{\rm K}^{+}} \, n_{\rm
    e} / n_{{\rm K}^{0}}$, as a function of temperature, for various
  values of the work function, $W$.  The value of $W$ (in eV) is used
  to label several of the curves.  The case $W = 5.0 \, {\rm eV}$ is
  bolded.  The horizontal dashed line refers to $E = 4.34 \, {\rm
    eV}$, the value tht would obtain in the absence of species other
  than potassium.  Because of emission of charges from dust grains
  (thermionic emission of electrons at low $W$, ion emission of ${\rm
    K}^{+}$ at large $W$), the gas charge state is always greater than
  expected based on a pure Saha equation for potassium; that is, $E <
  4.34 \, {\rm eV}$.  The effect is most pronounced and sensitive to
  $W$ at low temperatures $\ltsimeq \, 800 \, {\rm K}$.  For $T \,
  \gtsimeq \, 800 \, {\rm K}$, $3.5 {\rm eV} \, \ltsimeq \, E \,
  \ltsimeq \, 4.34 \, {\rm eV}$ for $W$ between 2 and 6~eV.  For work
  functions close to the ionization potential of potassium ($W \approx
  4.34 \, {\rm eV}$), $E \approx 4.34 \, {\rm eV}$.}
\end{center}
\label{fig:ECanonical}
\end{figure}
%
%-------------------------------------------------------------------------------

%Figure~\ref{fig:tCanonical} 
Figure~5 shows the timescales for the steady-state electron density
$n_{\rm e}$ to be affected by various ionization / recombination
processes, as functions of temperature.  For example, since ${\cal
  R}_{\rm therm}$ is the rate per volume of nebula at which electrons
are produced, $t_{\rm therm} = n_{\rm e} / {\cal R}_{\rm therm}$.
Likewise, the timescale for recombination of electrons with atomic
ions is defined as $n_{\rm e} / (\beta \, n_{\rm e} \, n_{\rm i}) = 1
/ (\beta \, n_{\rm i})$.  At low temperatures, $T < 700 \, {\rm K}$,
the dominant loss mechanism is adsorption of electrons onto dust
grains, while the dominant production mechanism is ionizations of
gas-phase molecules by energetic particles.  These balance each other
and the electron density turns over on timescales of roughly
10~seconds.  At temperatures above 800~K, thermionic emission of
electrons from grains dominates, and the electron density increases.
Ionizations by energetic particles become less important compared to
thermionic emission.  But because the electron density is greater, the
timescale for overturning the electron population rises, to above
1~hour at about 780~K, to about 1~day at 870~K, and peaking at
$10^6$~s at 900~K.  At higher temperatures still, the timescales for
both adsorption and thermionic emission decrease, eventually falling
below about 1~hour again for $T \, \gtsimeq \, 1300 \, {\rm K}$.
%-------------------------------------------------------------------------------
%
% FIGURE 5 
%
\begin{figure}[H]
\begin{center} 
\includegraphics[width=0.80\paperwidth]{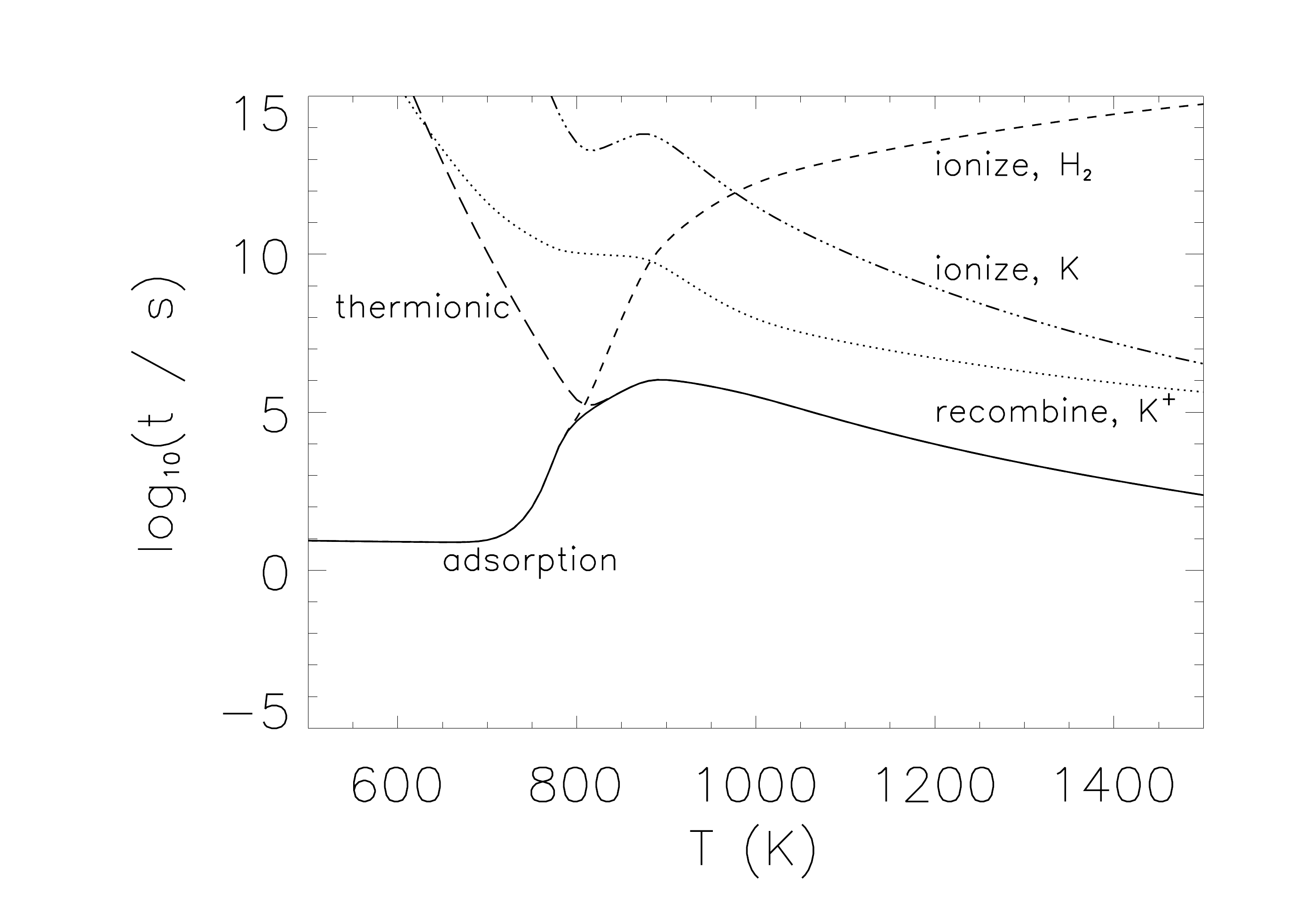}
\caption{Characterisic timescales for various processes affecting the
  electron abundances, as functions of temperature, for our canonical
  case.  The dominant removal processes is always electrons colliding
  with grains (`adsorption', solid curve).  At low temperatures ($T \,
  \ltsimeq \, 800 \, {\rm K}$), ionization of ${\rm H}_{2}$ by
  energetic particles ('ionize, ${\rm H}_{2}$', dashed curve) is the
  dominant charging process.  At higher temperatures, emission of
  electrons from grains ('thermionic', long dashed curve) dominates.
  At high temperatures the timescales for thermal ionization of K
  (dashed-dotted curve) and recombinations of ${\rm K}^{+}$ ions and
  electrons (dotted curve) become shorter but never dominate.
  Recombination of atomic ions and electrons is slower than all other
  processes and is not plotted.  Notably, for temperatures $800 \,
  {\rm K} \, \ltsimeq \, T \, \ltsimeq \, 1300 \, {\rm K}$, the
  chemical timescales are longer than hours.}
\end{center}
\label{fig:tCanonical}
\end{figure}
%
%-------------------------------------------------------------------------------

\subsection{Variation with Work Function}

We now investigate the effect of varying the work function of the
solids making up the grains, keeping other parameters the same as our
canonical case ($n_{{\rm H}_{2}} = 10^{14} \, {\rm cm}^{-3}$,
dust-to-gas ratio $10^{-2}$).
% Figure~\ref{fig:ZCanonical} 
Figure~3 shows the charge on each dust grain (in units of $e$) as a
function of temperature $T$ for various $W$.  At low work function, $W
\, \ltsimeq \, 4 \, {\rm eV}$, thermionic emission of electrons is
very effective, whereas any potassium atoms leaving are likely to
become neutral ${\rm K}^{0}$ atoms.  The grains are positively
charged.  At higher temperatures, the rate of thermionic emission
is greater, and the dust grains are increasingly positively charged.
For high work function, $W \, \gtsimeq \, 4 \, {\rm eV}$, thermionic
emission is ineffective, but emission of ${\rm K}^{+}$ is effective.
At higher temperatures, grains emit ions more effectively and so are
more negatively charged.  At temperatures $\approx 800$ to 1000~K,
however, ion emission is as effective as it can be.  At still higher
temperatures, gas-phase ions' collisions with dust grains are
increasingly important, and the grains become less negatively charged.

% Figure~\ref{fig:ECanonical} 
Figure~4 shows the effective ionization potential, $E$, needed for the
Saha equation to yield our calculated value of $n_{{\rm K}^{+}} \,
n_{\rm e} / n_{{\rm K}^{0}}$, as a function of temperature, for
various work functions $W$.  At low temperatures this ratio depends
sensitively on $W$, as the ion and electron abundances are controlled
by grain processes.  Especially for high work functions, $W \,
\gtsimeq \, 4 \, {\rm eV}$, ion emission dominates and increases
$n_{{\rm K}^{+}}$ above the levels predicted using the Saha equation.
At low work functions, thermionic emission increases $n_{\rm e}$.  In
gas hotter than about 800~K, thermal ionizations dominate and $E$ is
driven toward the ionization potential ${\rm IP} = 4.34 \, {\rm eV}$;
however, this is only a trend, and the effective ionization potential
falls in the range $3.5 \, {\rm eV} \, \ltsimeq \, E \, \ltsimeq \,
4.34 \, {\rm eV}$.  The departure of $E$ from ${\rm IP}$ is most
pronounced for $W$ that are far from ${\rm IP}$; for $W \approx 4.34
\, {\rm eV}$, $E \approx 4.34 \, {\rm eV}$.  Notably, though, $E$
never appreciably exceeds ${\rm IP}$, meaning the dust makes the gas
better-ionized than would be predicted using the Saha equation alone.

% Figure~\ref{fig:xeWsix} 
Figure~6 shows the abundances of various species if the work function
is increased to $W = 6.0 \, {\rm eV}$.  Compared to the case with $W =
5.0 \, {\rm eV}$, the abundances as a function of temperature $T$ are
quite similar, with the notable differences that $n_{{\rm K}^{0}}$ is
much reduced, and $n_{{\rm K}^{+}}$ rises with $T$ much more steeply.
This underscores that what is driving up $n_{{\rm K}^{+}}$ for $T \,
\ltsimeq \, 800 \, {\rm K}$ is ions evaporating from grains, and
whether the K atoms leave as ions or neutral atoms is a strong
function of $W$, or more precisely ${W} - {\rm IP}$.  Because of the
grains' large negative charge, atomic ions' abundances are suppressed.
%-------------------------------------------------------------------------------
%
% FIGURE 6 
%
\begin{figure}[H]
\begin{center} 
\includegraphics[width=0.80\paperwidth]{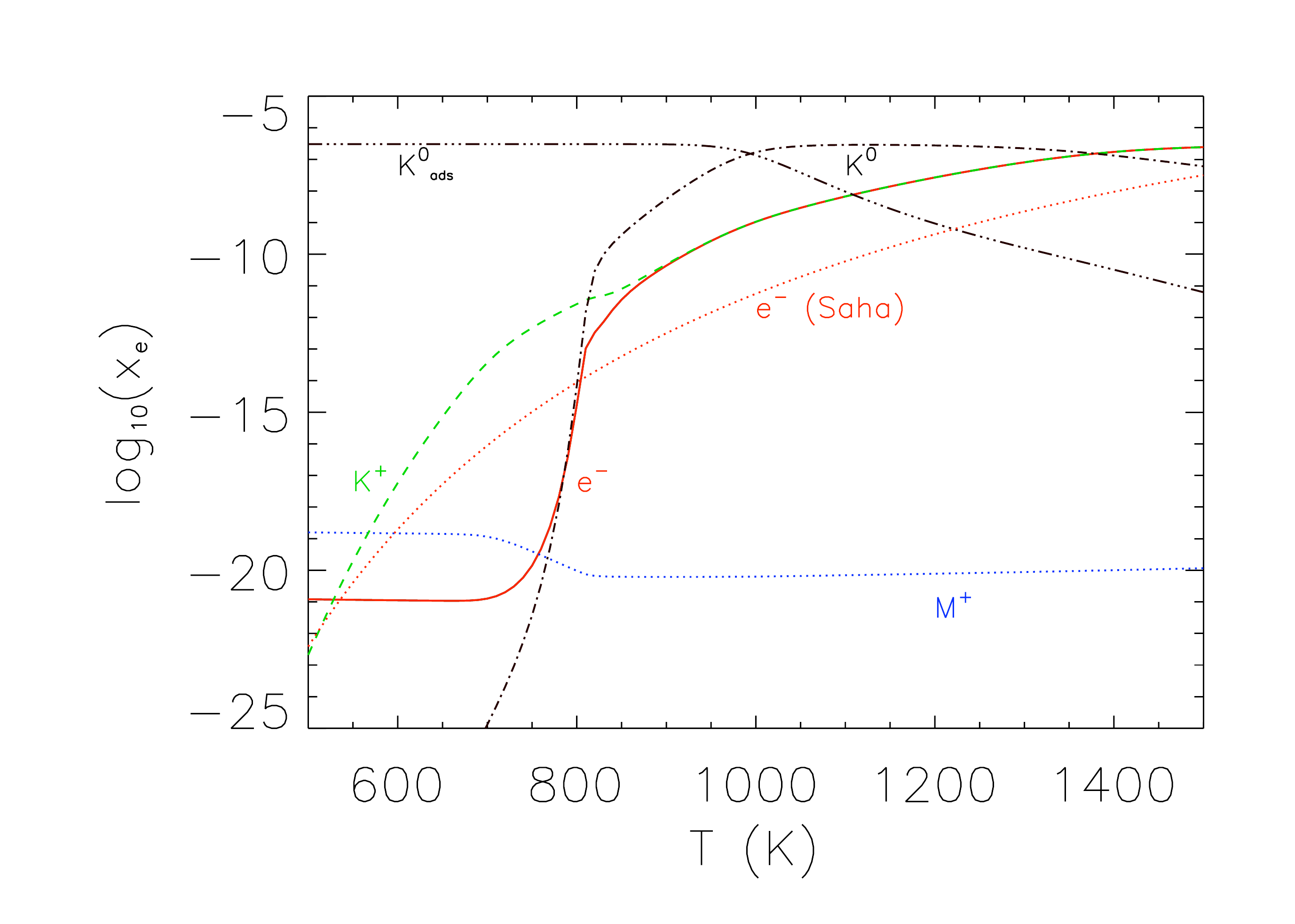}
\caption{Abundances of electrons (solid red curve), atomic (Mg) ions
  (dotted blue curve), ${\rm K}^{+}$ ions (dashed green curve), gas-phase 
  ${\rm K}^{0}$ atoms (dash-dot curve), and bound K atoms
  (dash-dot-dot-dot curve), as functions of temperature $T$, for the
  case with canonical parameters (see Figure 2), but with grain work
  function $W = 6.0 \, {\rm eV}$.  The electron fraction predicted using 
  the Saha equation is shown for comparison (dashed red curve). }
\end{center}
\label{fig:xeWsix} 
\end{figure}
%
%-------------------------------------------------------------------------------

% Figure~\ref{fig:xeWtwo} 
Figure~7 shows the abundances if the work function is decreased to $W
= 2.0 \, {\rm eV}$.  Again, the general behavior with temperature is
similar, except for $T \, \ltsimeq \, 800 \, {\rm K}$.  For the $W =
2.0 \, {\rm eV}$ case, production of ${\rm K}^{+}$ ions is
considerably lower; according to the Saha-Langmuir equation, nearly
all K atoms leave as neutral atoms.  On the other hand, production of
free electrons by thermionic emission is much more effective.  The
electron fraction rises to $x_{\rm e} \sim 10^{-12}$ and plateaus
there for temperatures between about 550~K and 900~K.  Because of the
large positive charge on each grain, the gas-phase ions' abundances
are elevated.
%-------------------------------------------------------------------------------
%
% FIGURE 7 
%
\begin{figure}[H]
\begin{center} 
\includegraphics[width=0.80\paperwidth]{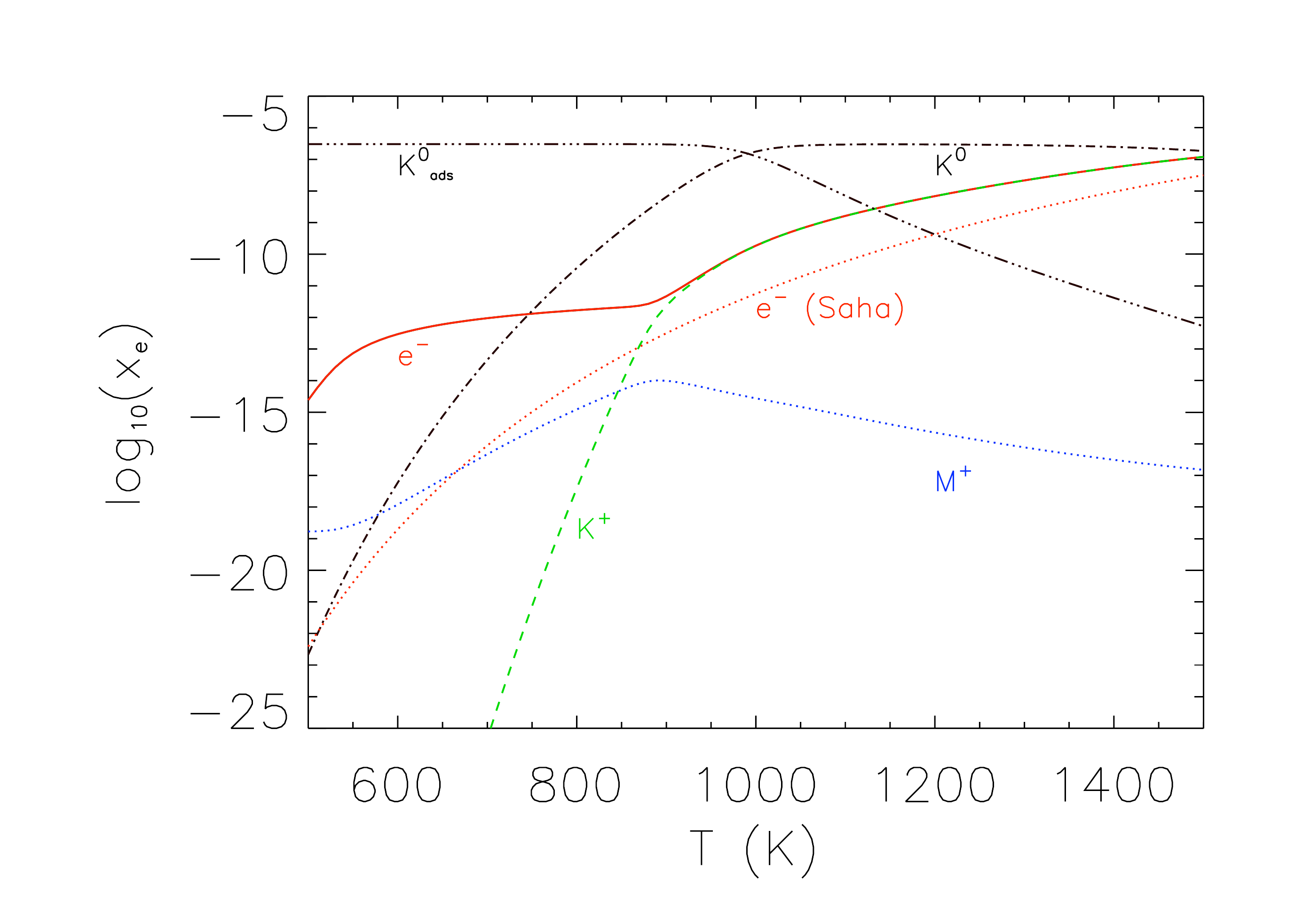}
\caption{Abundances of electrons (solid red curve), atomic (Mg) ions
  (dotted blue curve), ${\rm K}^{+}$ ions (dashed green curve), gas-phase 
  ${\rm K}^{0}$ atoms (dash-dot curve), and bound K atoms
  (dash-dot-dot-dot curve), as functions of temperature $T$, for the
  case with canonical parameters (see Figure 2), but with grain work
  function $W = 2.0 \, {\rm eV}$.  The electron fraction predicted using 
  the Saha equation is shown for comparison (dotted red curve). }
\end{center}
\label{fig:xeWtwo} 
\end{figure}
%
%-------------------------------------------------------------------------------

Finally, % Figure~\ref{fig:xeallW} 
Figure~8 shows just the electron abundance as a function of
temperature, for various values of the grain work function.  At high
temperatures, $\gtsimeq \, 1000 \, {\rm K}$, the behavior of $x_{\rm
  e}$ with $T$ is very similar, but the exact value does depend on the
work function, although not monotonically with $W$.  At lower
temperatures the electron fraction is especially sensitive to the work
function, being up to 8 orders of magnitude higher for low values ($W
\approx 2 \, {\rm eV}$) than for high values ($W \approx 6 \, {\rm
  eV}$).
%-------------------------------------------------------------------------------
%
% FIGURE 8 
%
\begin{figure}[H]
\begin{center} 
\includegraphics[width=0.80\paperwidth]{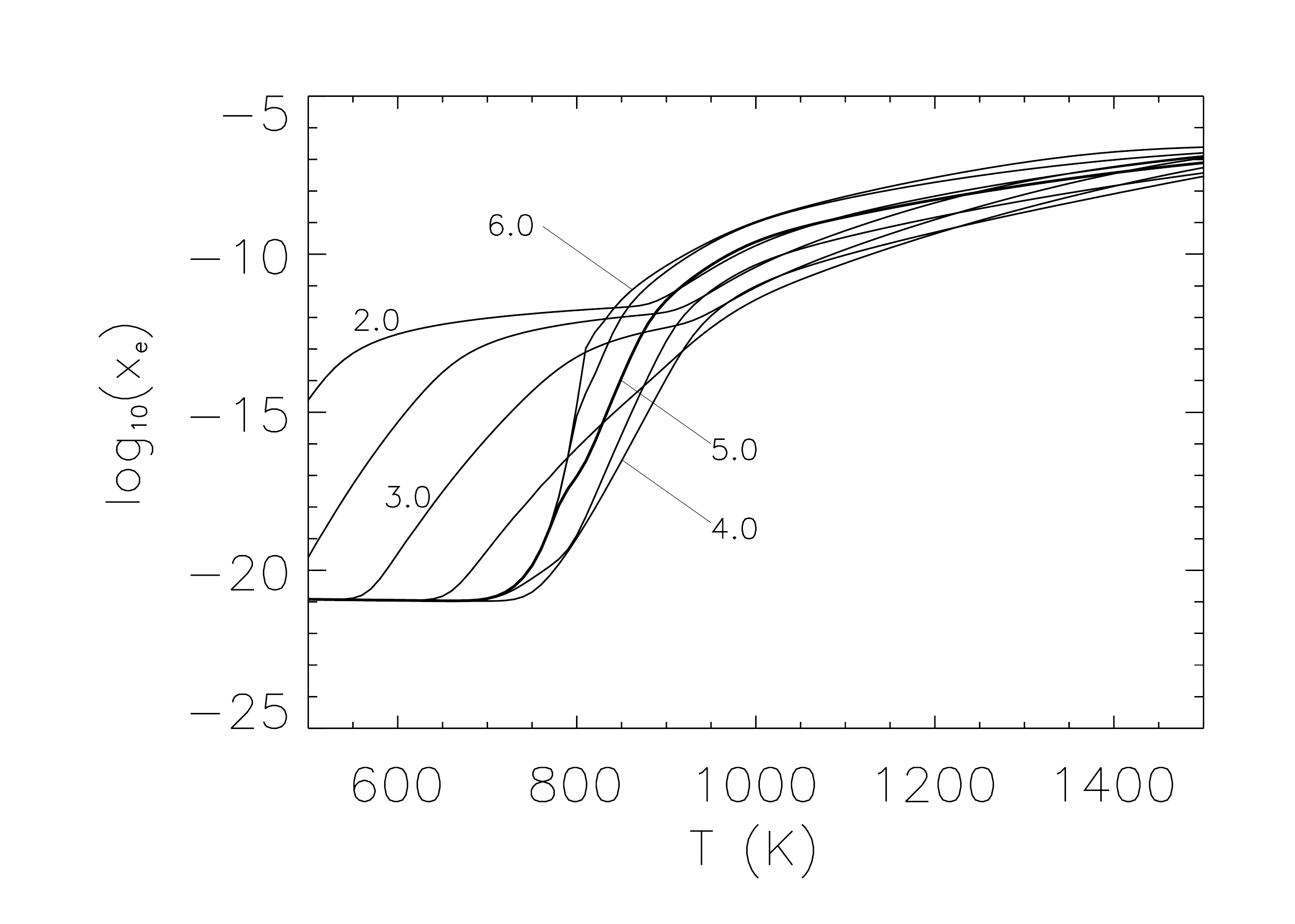}
\caption{Electron fraction as a function of temperature, for various
  grain work functions.  Other parameters are held at their values in
  the canonical case.  The fiducial case ($W = 5.0 \, {\rm eV}$) is in
  bold.  Electron fractions at low temperature ($\ltsimeq \, 800 \,
  {\rm K}$) decrease as the work function is increased from $2.0 \,
  {\rm eV}$ to $3.0 \, {\rm eV}$ to $4.0 \, {\rm eV}$ (as labeled).
  As the work function is increased more, from 4 to 6~eV, the electron
  fraction tends to increase for temperatures $T \approx 800 - 900 \,
  {\rm K}$.}
\end{center}
\label{fig:xeallW} 
\end{figure}
%
%-------------------------------------------------------------------------------

\subsection{Variation with Dust-to-Gas Ratio}

We now investigate the effect of varying the dust-to-gas ratio,
keeping other parameters the same as our canonical case ($W = 5 \,
{\rm eV}$, $n_{{\rm H}_{2}} = 10^{14} \, {\rm cm}^{-3}$).
% Figure~\ref{fig:xeDfour} 
Figure~9 shows the abundances of various species for the case of a
dust-to-gas ratio 0.0001, a factor 100 below the canonical value.
% Figure~\ref{fig:xeDone} 
Figure~10 shows the abundances of various species for a dust-to-gas
ratio of unity, a factor 100 higher than the canonical value.
Comparing % Figures~\ref{fig:xeDfour} and~\ref{fig:xeCanonical},
Figures~9 and 2, for the reduced dust-to-gas ratio both the electron
and ion densities are higher at low temperatures ($T < 600$ K),
because both are produced by energetic particles and lost from the gas
phase by adsorption onto grains.  Comparing
%Figures~\ref{fig:xeDone} and~\ref{fig:xeCanonical},
Figures~10 and 2, for the increased dust-to-gas ratio these densities
are reduced below the canonical case.  At high temperatures ($T \,
\gtsimeq \, 1000$ K), all three dust-to-gas ratios lead to similar
electron densities, and $n_{\rm e} \approx n_{{\rm K}^{+}}$, because
in that regime thermal ionizations of potassium dominate.  At
intermediate temperatures, there is a strong tendency for the electron
and ion (both atomic ions and ${\rm K}^{+}$) densities to decrease
with increasing dust-to-gas ratio, which makes sense given that they
are lost mostly to collisions with dust grains.
%-------------------------------------------------------------------------------
%
% FIGURE 9 
%
\begin{figure}[H]
\begin{center} 
\includegraphics[width=0.80\paperwidth]{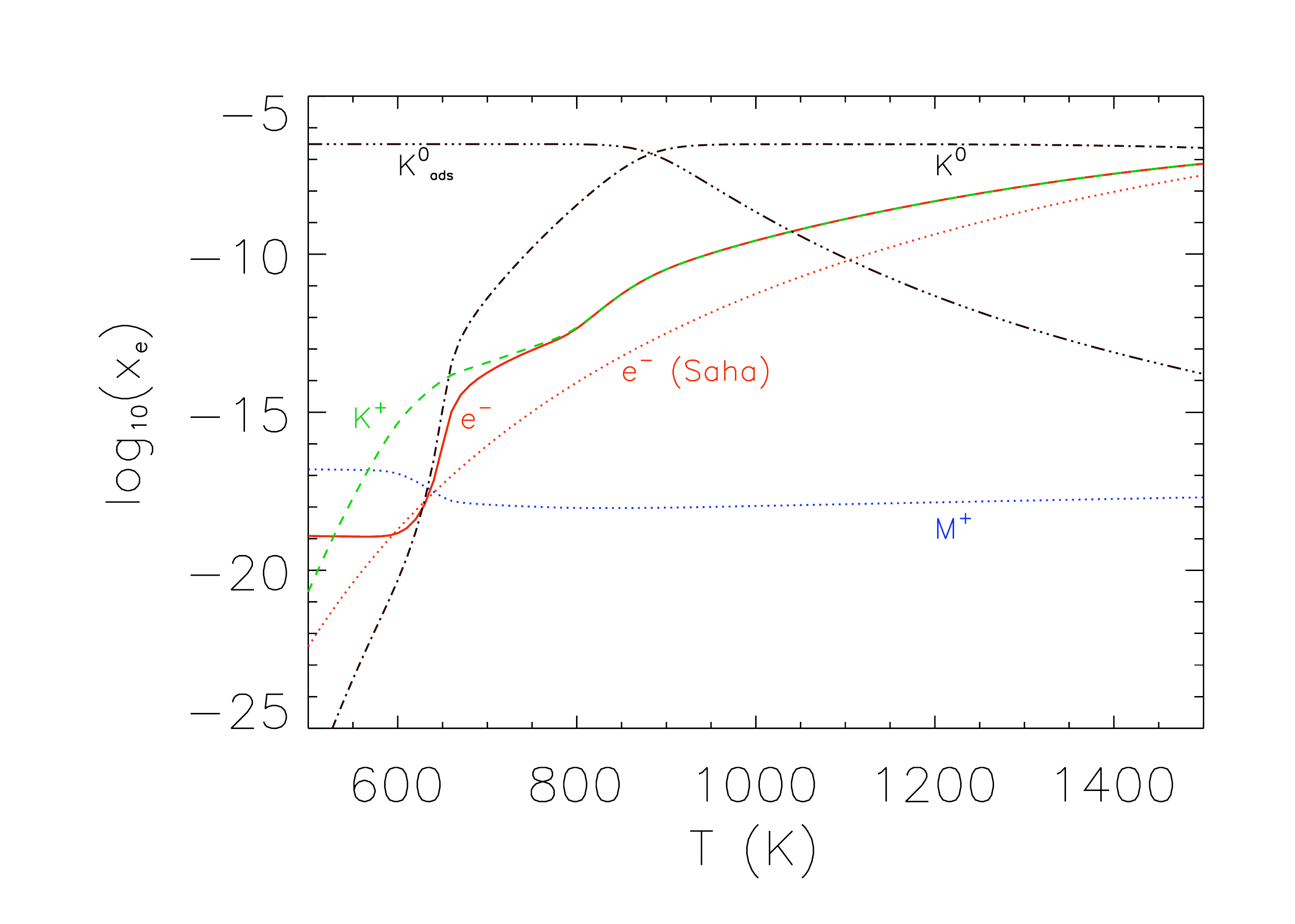}
\caption{Abundances of electrons (solid curve), atomic (Mg) ions
  (dotted curve), ${\rm K}^{+}$ ions (dashed curve), gas-phase ${\rm
    K}^{0}$ atoms (dash-dot curve), and bound K atoms
  (dash-dot-dot-dot curve), as functions of temperature $T$, for the
  case with canonical parameters (see Figure 2), but with dust-to-gas
  mass ratio $10^{-4}$.
  The electron fraction predicted using the Saha equation (dotted red curve)
  is shown for comparison. }
\end{center}
\label{fig:xeDfour} 
\end{figure}
%
%-------------------------------------------------------------------------------

%-------------------------------------------------------------------------------
%
% FIGURE 10
%
\begin{figure}[H]
\begin{center} 
\includegraphics[width=0.80\paperwidth]{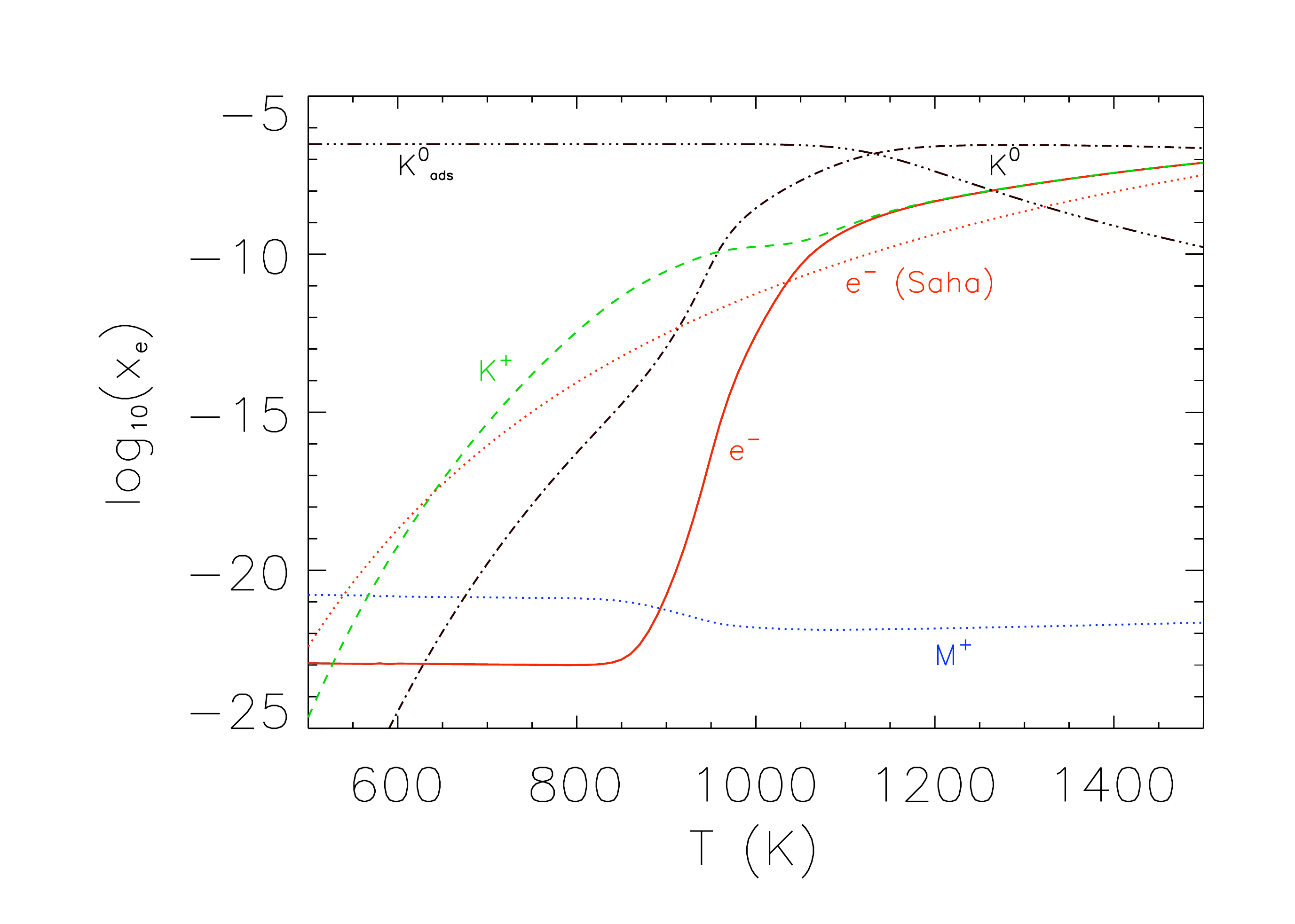}
\caption{Abundances of electrons (solid curve), atomic (Mg) ions
  (dotted curve), ${\rm K}^{+}$ ions (dashed curve), gas-phase ${\rm
    K}^{0}$ atoms (dash-dot curve), and bound K atoms
  (dash-dot-dot-dot curve), as functions of temperature $T$, for the
  case with canonical parameters (see Figure 2), but with dust-to-gas
  mass ratio $1$.
  The electron fraction predicted using the Saha equation (dotted red curve)
  is shown for comparison. 
}
\end{center}
\label{fig:xeDone} 
\end{figure}
%
%-------------------------------------------------------------------------------

% Figure~\ref{fig:ZallD} 
Figure~11 shows the grain charge as a function of temperature, for
various dust-to-gas ratios between $10^{-4}$ and unity.  For all these
cases, the charge on each grain is small at low temperatures, but
becomes sharply negative as the temperature increases above some
threshold; for the canonical ratio $10^{-2}$ that threshold
temperature is 700~K.  As the temperature is increased further, the
grain attains a maximum negative charge but then becomes less charged,
eventually (above 1000~K in the canonical case) following $Z \approx
-500 \, (T / 1200 \, {\rm K})^{-1}$ in all cases.  For other
dust-to-gas ratios the behaviors are the same but the thresholds occur
at temperatures 100~K lower if the dust-to-gas ratio is $10^{-4}$, or
200~K higher if the dust-to-gas ratio is unity.

Finally, % Figure~\ref{fig:xeallD} 
Figure~12 shows the electron fraction as a function of temperature for
various values of the dust-to-gas ratio.  As might be expected,
$x_{\rm e}$ is inversely proportional to this ratio at low
temperatures ($\ltsimeq \, 600 \, {\rm K}$), and independent of this
ratio at high temperatures ($\gtsimeq \, 1100 \, {\rm K}$) where
gas-phase reactions dominate both ionization and recombination.  In
between, the electron fraction is highly sensitive to the dust-to-gas
ratio, being as much as 12 orders of magnitude higher for small values
($10^{-4}$), as compared to large values (unity).
%-------------------------------------------------------------------------------
%
% FIGURE 11 
%
\begin{figure}[H]
\begin{center} 
\includegraphics[width=0.80\paperwidth]{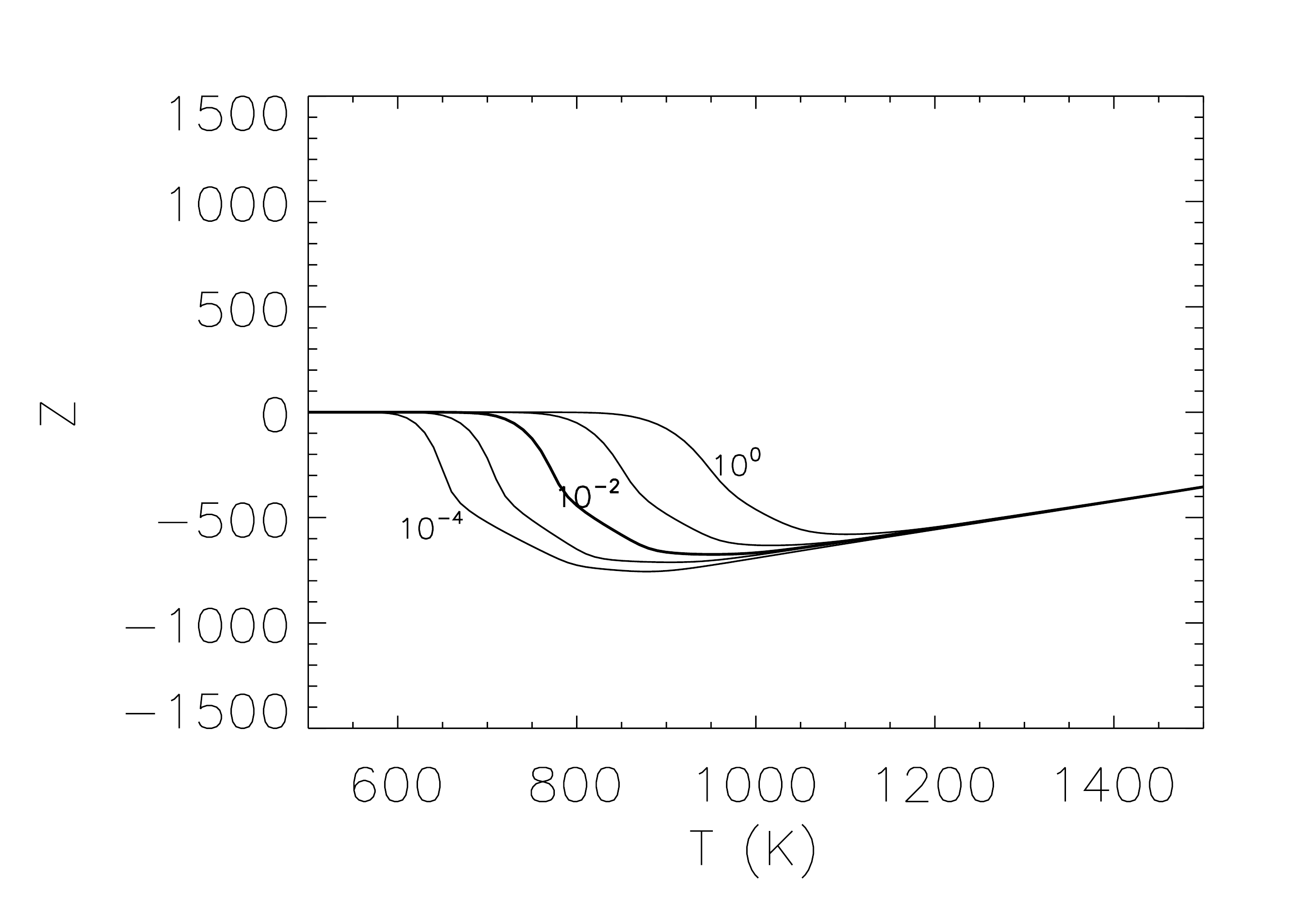}
\caption{Grain charge (in units of elementary charge $e$), as a
  function of temperature $T$, for various dust-to-gas ratios, from
  left to right $10^{-4}$, $10^{-3}$, $10^{-2}$, $10^{-1}$, and
  $10^{0}$.  The curve for the canonical dust-to-gas ratio, $10^{-2}$,
  is bolded.  With increasing temperature, the grains first become
  more negatively charged as they emit ${\rm K}^{+}$ ions, then become
  less negatively charged as they collide with ions.}
\end{center}
\label{fig:ZallD} 
\end{figure}
%
%-------------------------------------------------------------------------------
%-------------------------------------------------------------------------------
%
% FIGURE 12
%
\begin{figure}[H]
\begin{center} 
\includegraphics[width=0.80\paperwidth]{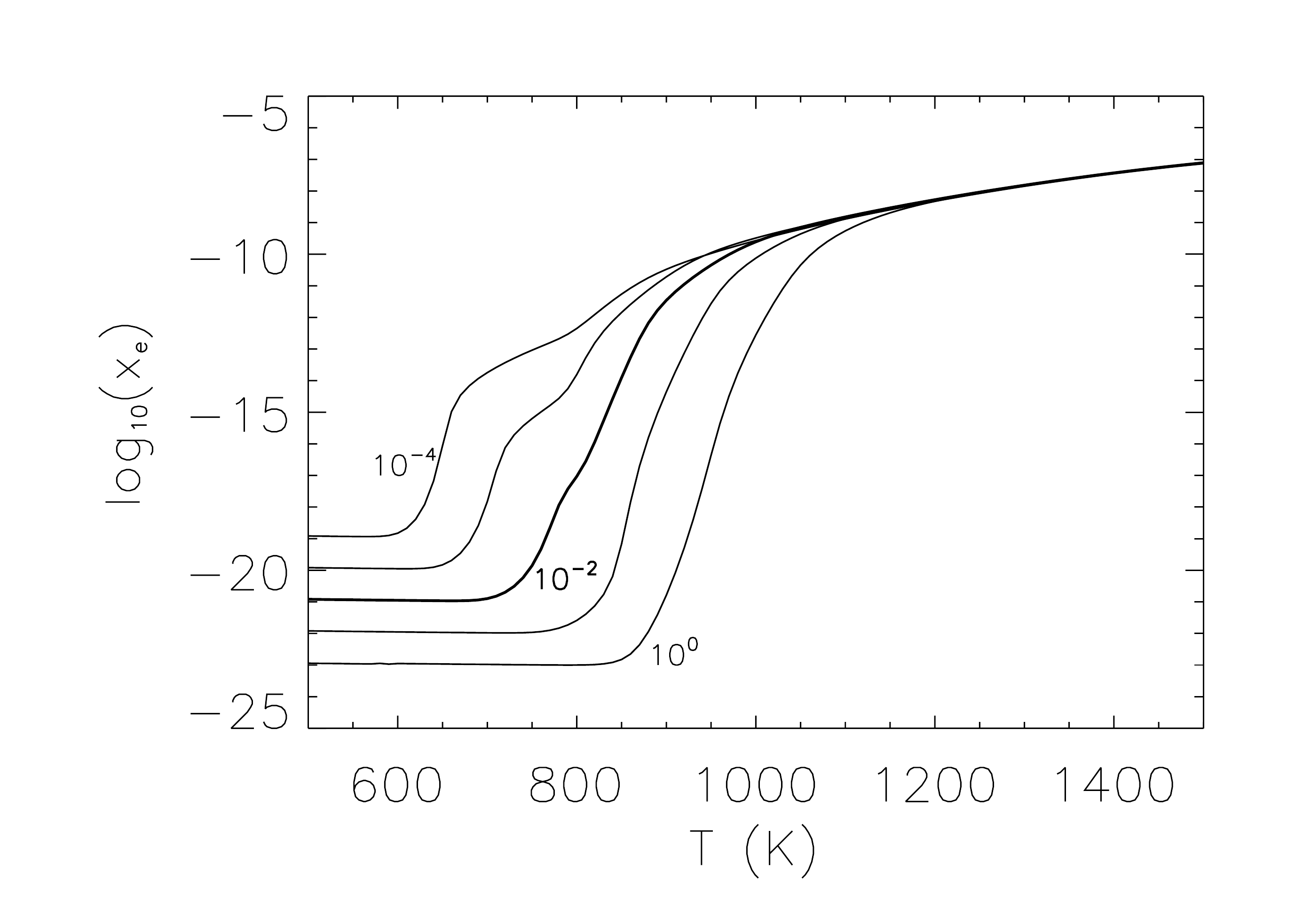}
\caption{Electron fraction vs.\ temperature for varying dust-to-gas
  ratios (as labeled).  Other parameters are held at their values in
  the canonical case. At low temperatures ($\ltsimeq \, 600 \, {\rm
    K}$), the electron fraction is inversely proportional to the
  dust-to-gas ratio.  At high temperatures ($\gtsimeq \, 1100 \, {\rm
    K}$, the electrons come mostly from thermal ionization and their
  abundance is independent of the dust-to-gas ratio.  At intermediate
  temperatures, the electron fraction is highly sensitive (but still
  inversely related) to the dust-to-gas ratio.}
\end{center}
\label{fig:xeallD} 
\end{figure}
%
%-------------------------------------------------------------------------------

\subsection{Variation with Gas Density}

We now investigate the effects of varying the gas density, keeping
other parameters the same as our canonical case ($W = 5 \, {\rm eV}$,
dust-to-gas ratio $10^{-2}$).  Figure~13 shows the electron fraction
as a function of temperature for the case of a gas density $n_{{\rm
    H}_{2}} = 10^{13} \, {\rm cm}^{-3}$, an order of magnitude below
the canonical value.  Figure~14 shows the electron fraction as a
function of temperature for the case of a gas density $n_{{\rm H}_{2}}
= 10^{16} \, {\rm cm}^{-3}$, a factor of 100 above the canonical
value.  Figure~15 shows the average charge on grains as a function of
temperature, for various gas densities.  The same trends are apparent:
at low temperatures ($\ltsimeq \, 700$~K), grains are nearly neutral,
becoming very negatively charged at temperatures of about 800 to
1000~K, then becoming less negatively charged at higher temperatures.
For higher gas densities, these transitions occur at slightly higher
temperatures; e.g., a ten-fold increase in $n_{{\rm H}_{2}}$ shifts
the transitions higher in temperature by only about 60~K.  Likewise,
Figure~16 shows the electron fraction as a function of gas density.
Again it is seen that the charging is relatively insensitive to the
gas density.
%-------------------------------------------------------------------------------
%
% FIGURE 13
%
\begin{figure}[H]
\begin{center} 
\includegraphics[width=0.80\paperwidth]{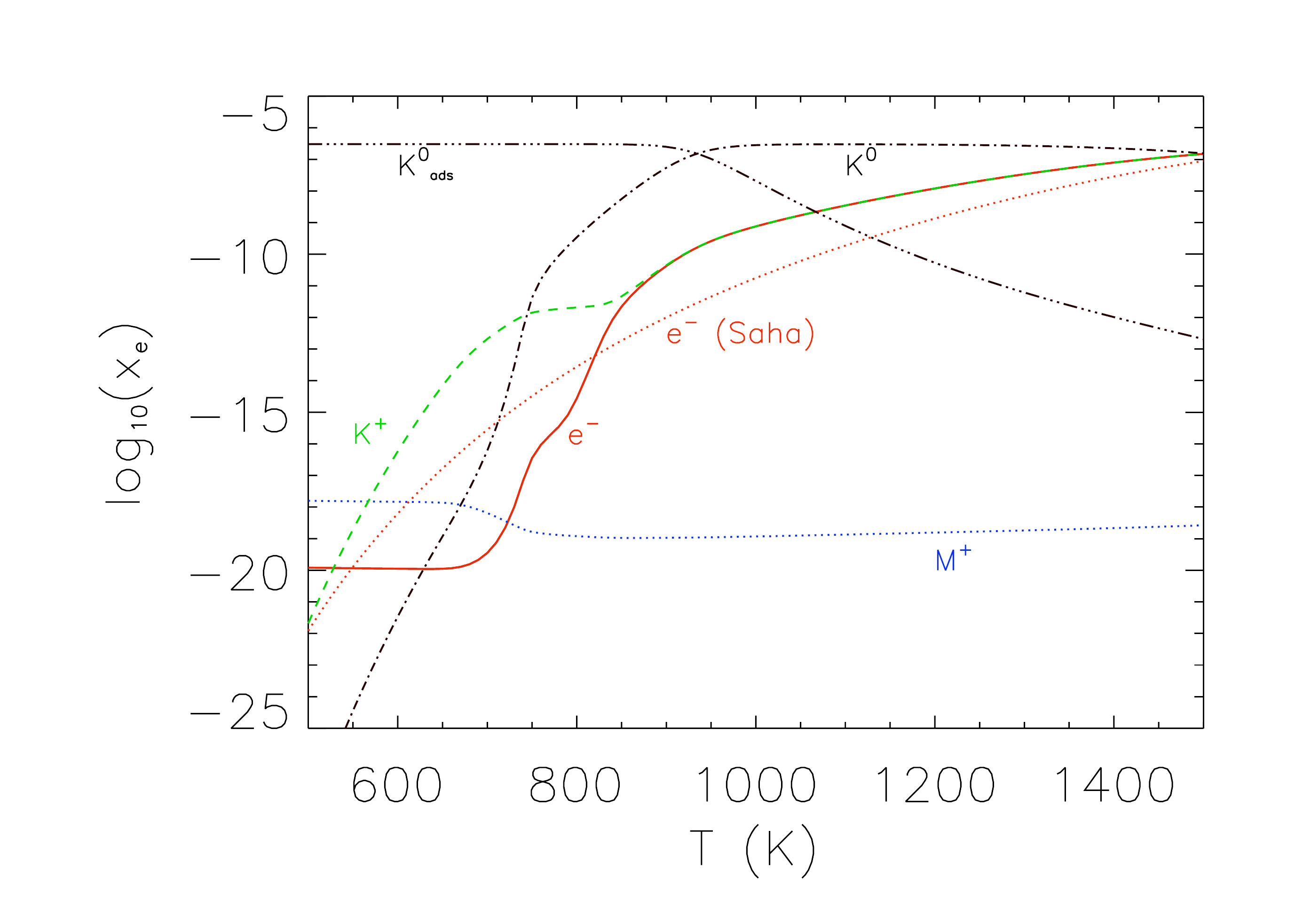}
\caption{Abundances of electrons (solid curve), atomic (Mg) ions
  (dotted curve), ${\rm K}^{+}$ ions (dashed curve), gas-phase ${\rm
    K}^{0}$ atoms (dash-dot curve), and bound K atoms
  (dash-dot-dot-dot curve), as functions of temperature $T$, for the
  case with canonical parameters (see Figure 2), but with gas density
  $10^{13} \, {\rm cm}^{-3}$.
 The electron fraction predicted using the Saha equation (red dotted curve)
  is shown for comparison. 
}
\end{center}
\label{fig:xeNthirteen}
\end{figure}
%
%-------------------------------------------------------------------------------
%-------------------------------------------------------------------------------
%
% FIGURE 14 
%
\begin{figure}[H]
\begin{center} 
\includegraphics[width=0.80\paperwidth]{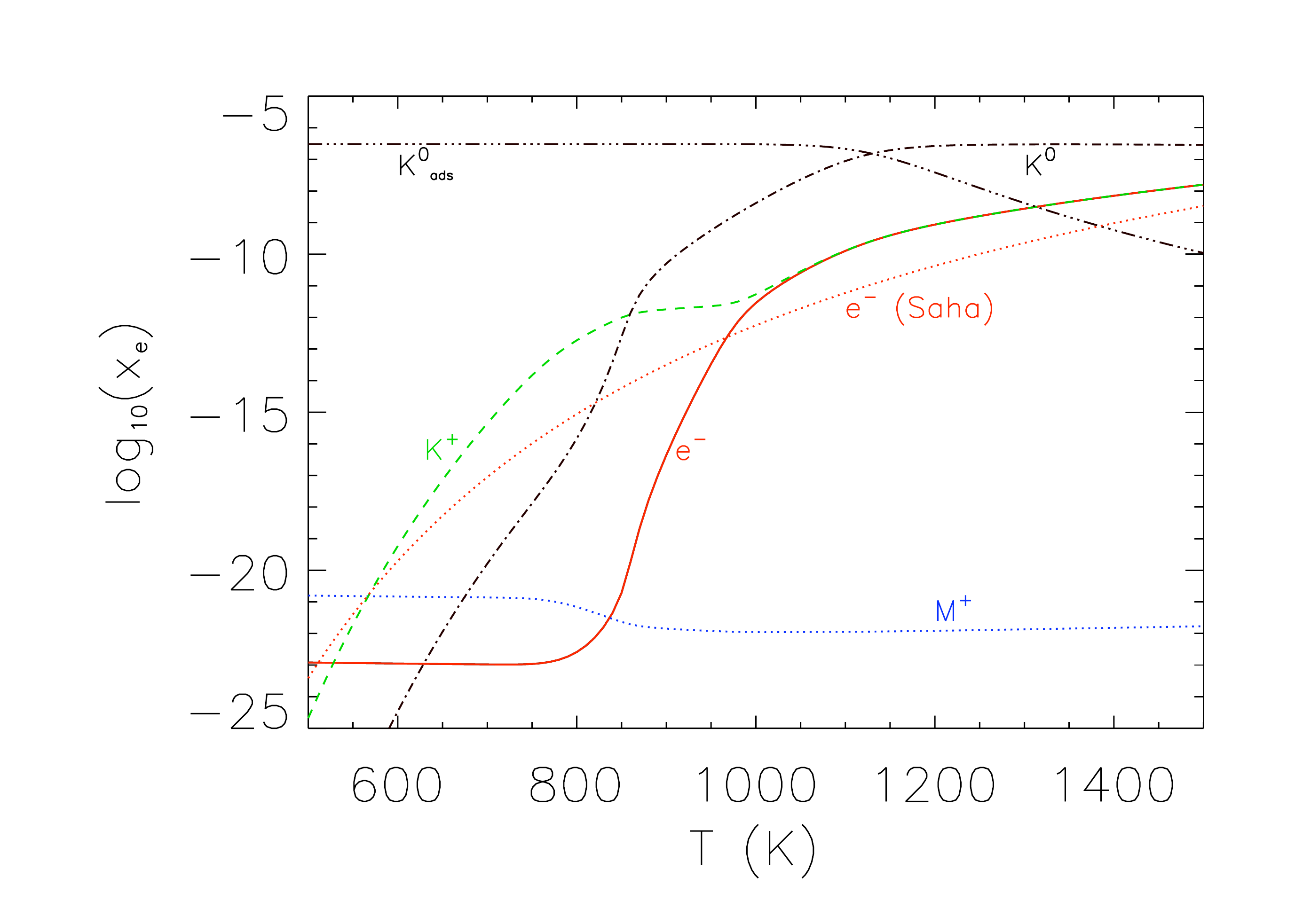}
\caption{Abundances of electrons (solid curve), atomic (Mg) ions
  (dotted curve), ${\rm K}^{+}$ ions (dashed curve), gas-phase ${\rm
    K}^{0}$ atoms (dash-dot curve), and bound K atoms
  (dash-dot-dot-dot curve), as functions of temperature $T$, for the
  case with canonical parameters (see Figure 2), but with gas density
  $10^{16} \, {\rm cm}^{-3}$.}
\end{center}
\label{fig:xeNsixteen}
\end{figure}
%
%-------------------------------------------------------------------------------
%-------------------------------------------------------------------------------
%
% FIGURE 15
%
\begin{figure}[H] 
\begin{center} 
\includegraphics[width=0.80\paperwidth]{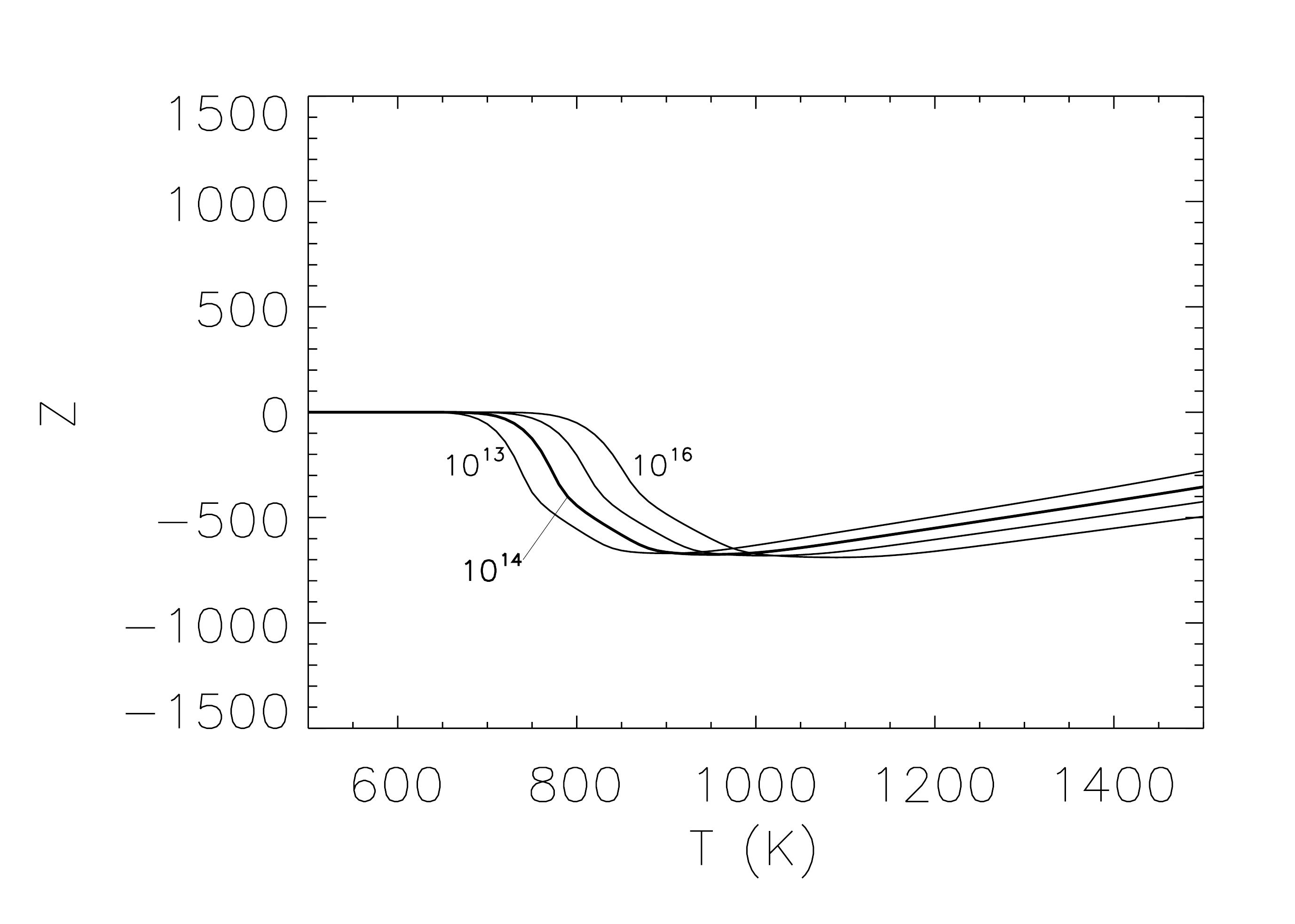}
\caption{Charge on dust grains as a function of temperature, for our canonical values
 (see Figure 2), but for various values of $n_{{\rm H}_{2}}$.}
\end{center}
\label{fig:ZallN}
\end{figure} 
%
%-------------------------------------------------------------------------------
%-------------------------------------------------------------------------------
%
% FIGURE 16
%
\begin{figure}[ht] 
\begin{center} 
\includegraphics[width=0.80\paperwidth]{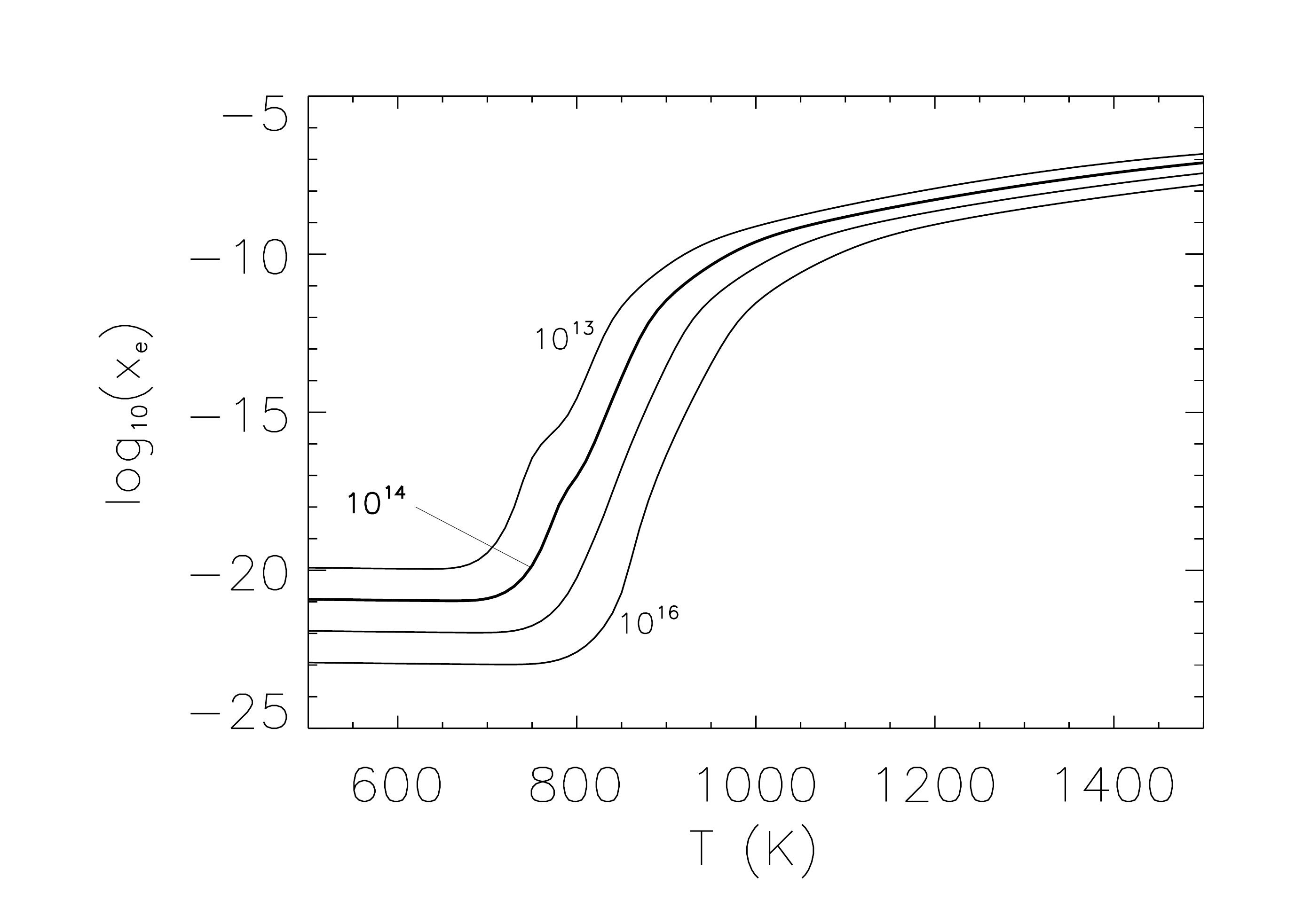}
\caption{Electron fraction as a function of temperature, for our canonical values
  (see Figure 2), but for various values of $n_{{\rm H}_{2}}$.}
\end{center}
\label{fig:xeallN}
\end{figure} 
%
%-------------------------------------------------------------------------------

\section{Application to Disks}

\subsection{Inner Edge of MRI Dead Zone}

Our chemical network allows us to calculate ionization levels in dusty
plasmas at high temperatures.  An important application is to
determining the inner edge of the MRI dead zone, that region of the
disk where the MRI cannot lead to turbulence at the midplane.  Jin
(1996) and Gammie (1996) both realized that ionization levels at PPD
midplanes are too low to permit good coupling of the gas to the
magnetic fields, at least in regions where the temperatures are below
about 1000~K.  Closer to the Sun, at heliocentric distances inside
roughly 0.1 to 1~AU, temperatures may exceed this threshold, and they
argued that thermal ionization of alkali elements in the gas would
lead to ionization levels high enough to permit the MRI.  Because the
mass flow changes abruptly at the dead zone edge, the location of the
transition bears significantly on the disk's structure and the
interpretation of its spectral energy distribution (e.g., Hasegawa \&
Pudritz 2010; Zhu et al.\ 2010).

\begin{figure}[ht] 
\begin{center} 
\includegraphics[width=0.80\paperwidth]{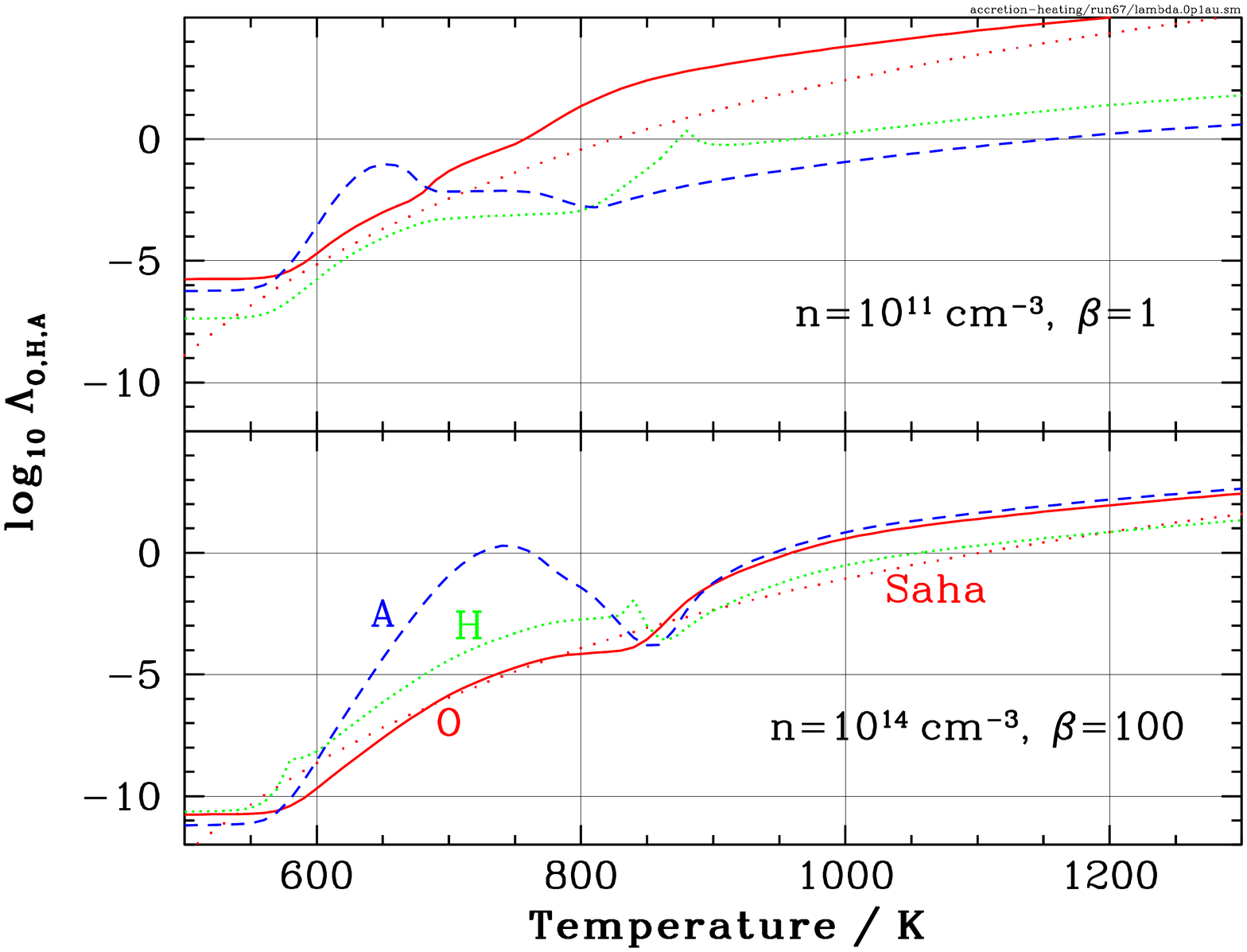}
\caption{Ohmic Elsasser number $\Lambda_{\rm O}$ (red solid curve),
  ambipolar Elsasser number $\Lambda_{\rm A}$ (blue dashed curve), and
  Hall Elsasser number $\Lambda_{\rm H}$ (green dotted curve) as
  functions of temperature at 0.1~AU from a Solar-mass star.  The red
  dotted curve shows the Ohmic Elsasser number computed from the Saha
  equation for potassium, neglecting thermionic and ion emission.
  Different values of the density and plasma beta parameter in the top
  and bottom panels (top: $n_{\rm H2} = 10^{11} \, {\rm cm}^{-3}$,
  $\beta = 1$; (bottom: $n_{\rm H2} = 10^{14} \, {\rm cm}^{-3}$,
  $\beta = 100$) correspond to the disk atmosphere and midplane,
  respectively.  \label{fig:elsasserone}}
\end{center}
\end{figure}

\begin{figure}[ht] 
\begin{center} 
\includegraphics[width=0.80\paperwidth]{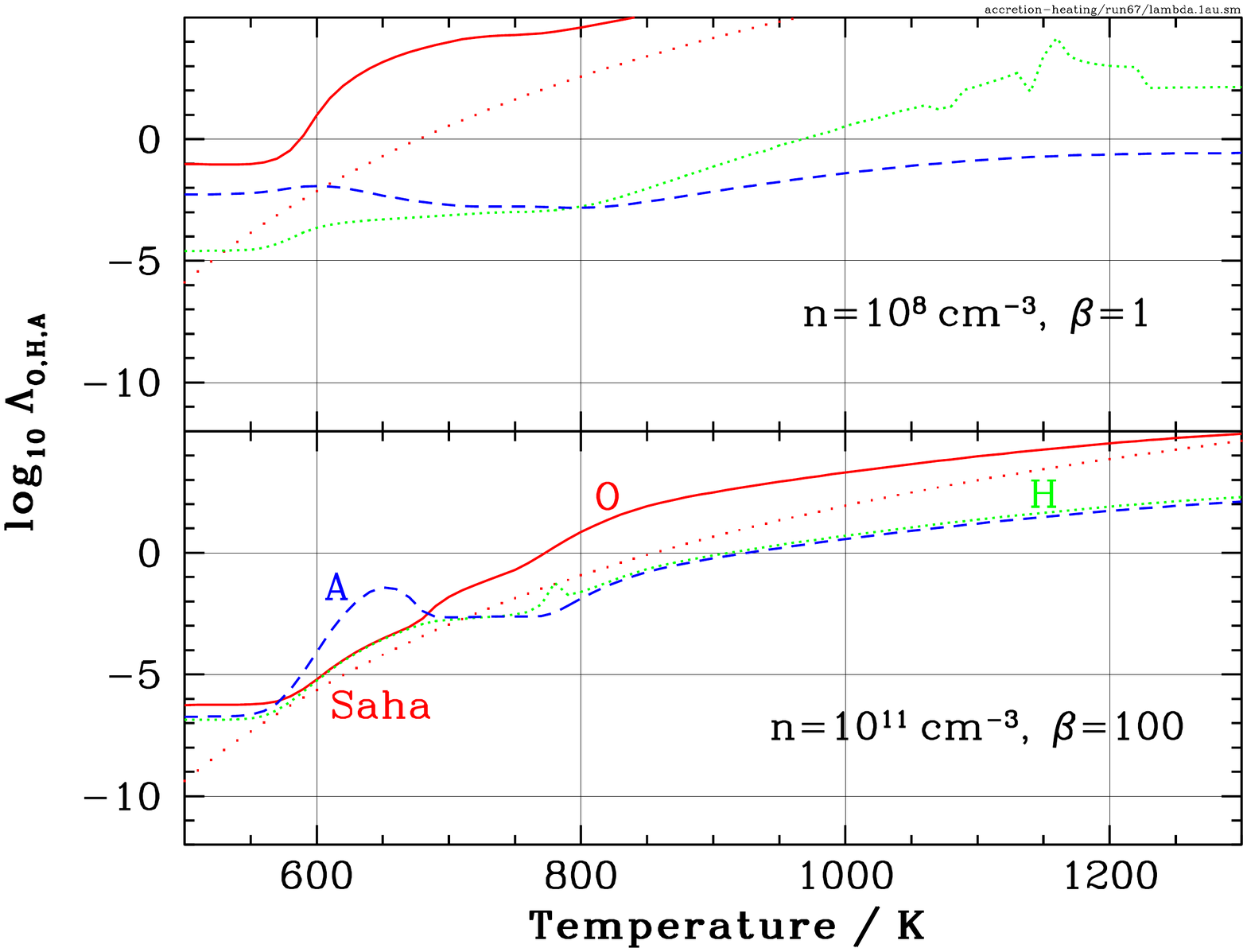}
\caption{Same as Figure~\ref{fig:elsasserone}, but for heliocentric
  distance 1~AU.  Ohmic Elsasser number $\Lambda_{\rm O}$ (red solid
  curve), ambipolar Elsasser number $\Lambda_{\rm A}$ (blue dashed
  curve), and Hall Elsasser number $\Lambda_{\rm H}$ (green dotted
  curve) are plotted vs.\ temperature $T$.  The red dotted curve shows
  the Ohmic Elsasser number computed from the Saha equation for
  potassium.  The top and bottom panels correspond to the disk
  atmosphere and midplane, respectively (top: $n_{\rm H2} = 10^{8} \,
  {\rm cm}^{-3}$, $\beta = 1$; bottom: $n_{\rm H2} = 10^{11} \, {\rm
    cm}^{-3}$, $\beta = 100$).
    \label{fig:elsassertwo}}
\end{center}
\end{figure}

\begin{figure}[ht] 
\begin{center} 
\includegraphics[width=0.80\paperwidth]{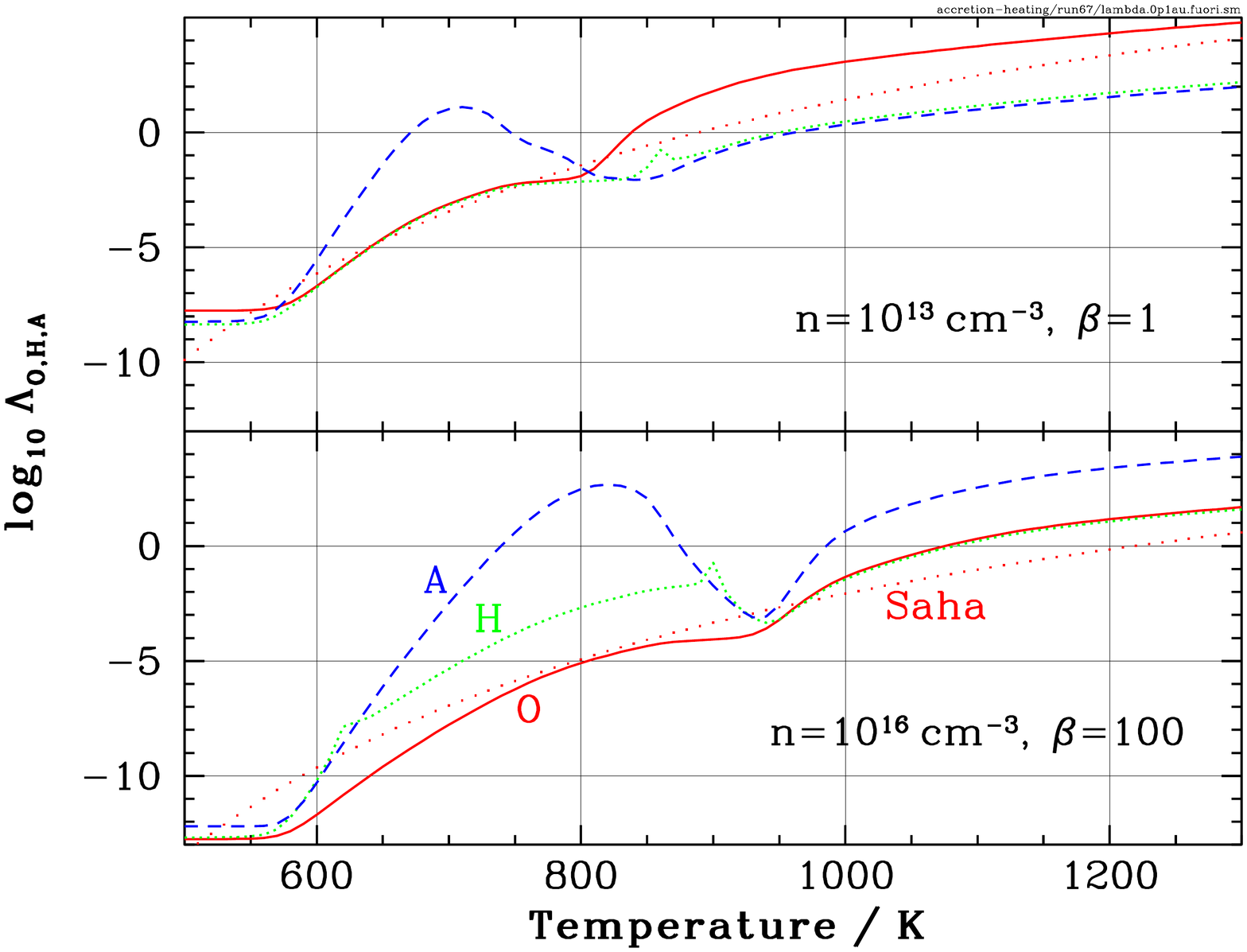}
\caption{Same as Figure~\ref{fig:elsasserone}, except that the
  densities and plasma beta parameters correspond to the disk
  atmosphere and midplane at the onset of an accretion outburst, when
  gas and dust have piled up near 0.1~AU (top panel: $n_{\rm H2} =
  10^{13} \, {\rm cm}^{-3}$, $\beta = 1$; bottom: $n_{\rm H2} =
  10^{16} \, {\rm cm}^{-3}$, $\beta = 100$).  Ohmic Elsasser number
  $\Lambda_{\rm O}$ (red solid curve), ambipolar Elsasser number
  $\Lambda_{\rm A}$ (blue dashed curve), and Hall Elsasser number
  $\Lambda_{\rm H}$ (green dotted curve) are plotted vs.\ temperature
  $T$.  The red dotted curve shows the Ohmic Elsasser number computed
  from the Saha equation for potassium.
  \label{fig:elsasserthree}}
\end{center}
\end{figure}

Using our chemical network, we identify precisely what temperatures
are required for sufficient coupling.  In
Figures~\ref{fig:elsasserone}-\ref{fig:elsasserthree} we plot the
Elsasser numbers associated with Ohmic dissipation ($\Lambda_{\rm
  O}$), ambipolar diffusion ($\Lambda_{\rm A}$) and the Hall effect
($\Lambda_{\rm H}$).  The Elsasser number $v_A^2/(\eta\Omega)$ is the
time in orbits for magnetic fields to diffuse across the
fastest-growing MRI wavelength; $v_{\rm A} = B / (4\pi \rho)^{1/2}$ is
the Alfven speed, $\Omega$ is the Keplerian orbital frequency, and
$\eta$ is the relevant diffusivity.  The Ohmic, ambipolar and Hall
diffusivities are straightforward to compute from the charged species'
abundances, gyrofrequencies around the magnetic field of strength $B$,
and rates of collision with other species (Desch 2004; Wardle 2007).
In Figures~\ref{fig:elsasserone}-\ref{fig:elsasserthree} we plot the
three Elsasser numbers as functions of temperature $T$, for various
combinations of heliocentric distance $r$, gas density $n_{\rm H2}$,
and magnetic field strength, set by fixing the plasma beta parameter
$\beta = P_{\rm gas} / (B^2 / 8\pi)$.  MRI turbulence is switched off
by an Ohmic Elsasser number $\Lambda_O<1$ (Sano \& Inutsuka 2001,
Turner et al.\ 2007) or an ambipolar Elsasser number $\Lambda_A<1$
(Bai \& Stone 2011), while Hall Elsasser numbers $\Lambda_H \ll 1$
replace the turbulence with a laminar flow in which magnetic torques
nevertheless yield rapid angular momentum transport (Kunz \& Lesur
2013; Lesur et al.\ 2014).

We assume a $1 \, M_{\odot}$ star in what follows.
Figure~\ref{fig:elsasserone} pertains to conditions at 0.1~AU.  The
top panel has $n_{\rm H2} = 10^{11} \, {\rm cm}^{-3}$ and $B \sim 1 \,
{\rm G}$, appropriate for the disk atmosphere well above the midplane,
and the bottom panel has $n_{\rm H2} = 10^{14} \, {\rm cm}^{-3}$ and
$B \sim 1 \, {\rm G}$, appropriate for the midplane.  As ionization
levels increase with increasing temperature, the Elsasser numbers also
generally increase.  For the disk atmosphere, the last quantity to do
so is clearly the one associated with ambipolar diffusion, which
therefore is the dominant mechanism suppressing the MRI and setting the
location of the dead zone inner edge.  In the disk atmosphere at 0.1
AU, temperatures must exceed 1150 K for the MRI to operate.  At the
midplane, ignoring the complications introduced by the Hall effect,
ambipolar and Ohmic diffusion are equally important for setting the
dead zone inner edge, which will lie where temperatures exceed about
950~K.

%-------------------------------------------------------------------------------
%
% FIGURE 17
%
%
%-------------------------------------------------------------------------------

Figure~\ref{fig:elsassertwo} pertains to conditions at 1~AU.  The top
panel has $n_{\rm H2} = 10^{8} \, {\rm cm}^{-3}$ and $B \sim 10^{-2}
\, {\rm G}$, appropriate for the disk atmosphere, and the bottom panel
has $n_{\rm H2} = 10^{11} \, {\rm cm}^{-3}$ and $B \sim 10^{-2} \,
{\rm G}$, appropriate for the midplane.  In the atmosphere, ambipolar
diffusion is again seen to be the main process suppressing the MRI,
and in the range of temperatures we consider, the MRI is never
allowed.  At the midplane, ambipolar diffusion once more dominates,
and initiating the MRI requires temperatures above about 900~K.

%-------------------------------------------------------------------------------
%
% FIGURE 18
%
%
%-------------------------------------------------------------------------------

Finally, Figure~\ref{fig:elsasserthree} pertains to conditions at
0.1~AU, but during an epoch of increased density, such as might occur
around the onset of an FU~Orionis accretion outburst.  The top panel
has $n_{\rm H2} = 10^{13} \, {\rm cm}^{-3}$ and $B \sim 10 \, {\rm
  G}$, appropriate for the disk atmosphere, and the bottom panel has
$n_{\rm H2} = 10^{16} \, {\rm cm}^{-3}$ and $B \sim 10 \, {\rm G}$,
appropriate for the midplane.  
In the atmosphere, ambipolar diffusion is again the dominant process 
suppressing the MRI; the Hall term is comparable.  
Temperatures must exceed about 950~K for the MRI to operate.  
At the midplane, Ohmic dissipation and the Hall effect are of similar strengths,
and temperatures must exceed about 1070~K for the MRI to operate.

%-------------------------------------------------------------------------------
%
% FIGURE 19
%
%
%-------------------------------------------------------------------------------

If these examples are representative, then it would seem that the dead zone's inner 
edge is set by the action of ambipolar diffusion, not Ohmic dissipation.  
In other respects, however, our results are similar to those that have been found 
using the Saha equation.  
In Figures~\ref{fig:elsasserone}-\ref{fig:elsasserthree} we also plot the Ohmic
Elsasser number under gas-phase collisional ionization of potassium, 
with the resistivity obtained from the standard formula 
$\eta = 234 \, T^{1/2} \, x_{\rm e}^{-1}$ (Blaes \& Balbus 1994; Equation 41). 
Our chemical network with thermionic and ion emission yields Ohmic resistivities 
lower, and Elsasser numbers higher, by an order of magnitude or more at temperatures near 1000 K.
% The results differ significantly from those using our chemical network including 
% thermionic and ion emission, but only up to temperatures no greater than 900~K.  
For the range of conditions we consider, initiating the MRI in disk atmospheres or midplanes
generally requires temperatures of at least 900 to 1150~K, remarkably similar to the generally 
assumed value of about 1000~K.

\subsection{Stellar Magnetospheric Loading}

Young stars grow by accreting material from their disks.  The gas
flows to the star along the stellar magnetic field lines if the mass
accretion rate is too low for the incoming gas to crush the stellar
magnetosphere (Edwards et al.\ 1994; Bouvier et al.\ 2007).  The gas
could be loaded onto the stellar field either through the field lines'
diffusion into the disk (Miller \& Stone 1997), through magnetic
reconnection (Hirose et al.\ 1997), or through a Rayleigh-Taylor type
instability (Romanova et al.\ 2008).  Which scenario predominates, and
how quickly the mass is loaded by each of the three mechanisms,
depends on the magnetic diffusivity.  We suggest that the transition
from loading by dynamical instability to loading by magnetic diffusion
occurs when the diffusivity exceeds $h^2\Omega$, where $h$ is the disk
density scale height.  This is based on the finding from ideal-MHD
modeling that the plasma beta is about unity at the magnetosphere's
edge where loading occurs (Romanova et al.\ 2012).  Consequently the
scale on which magnetic and Coriolis forces balance, and
fastest-growing vertical wavelength of the magnetorotational
instability, will be about $h$.

\begin{figure}[ht] 
\begin{center} 
\includegraphics[width=0.50\paperwidth]{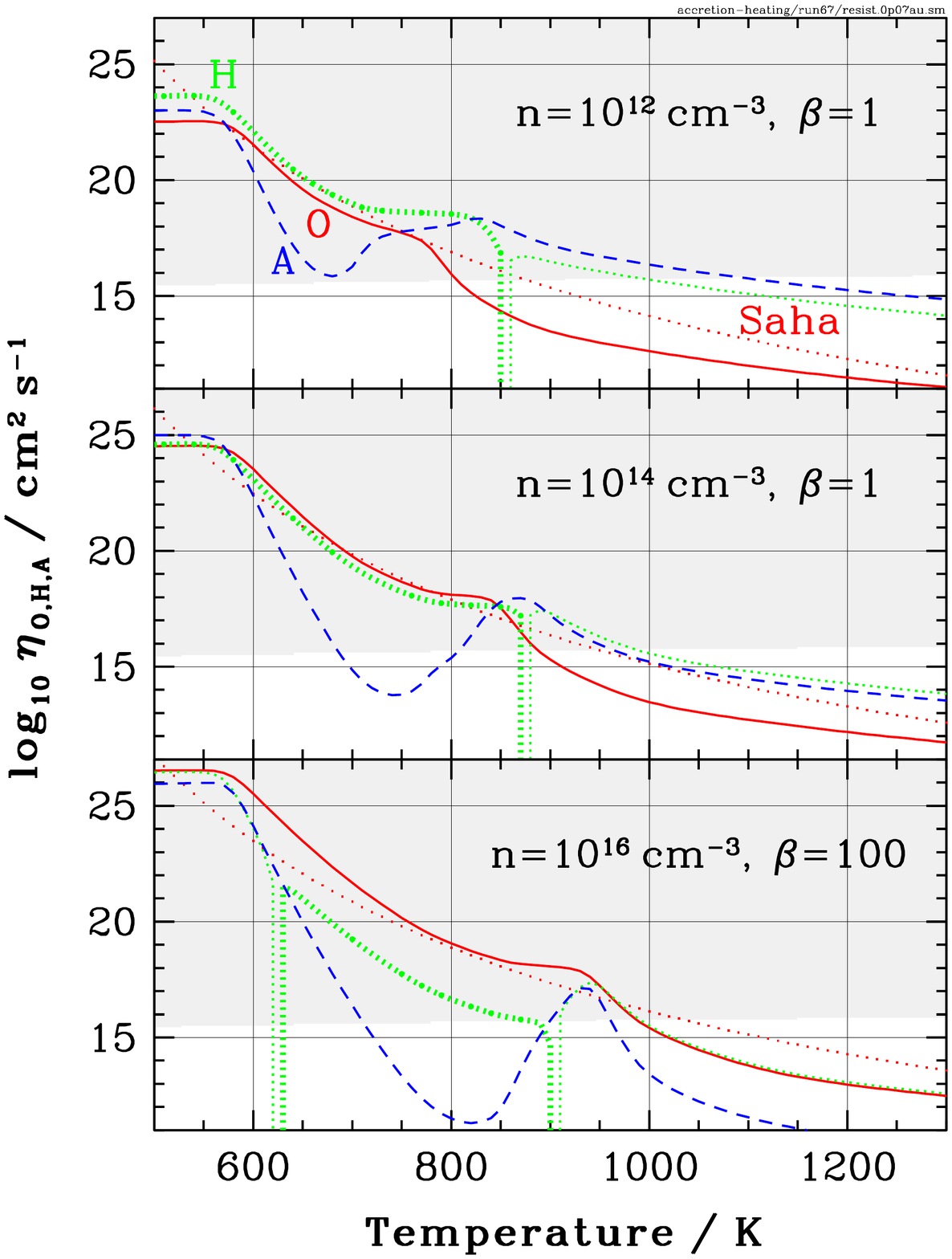}
\vspace{-0.3in} 
\caption{Magnetic diffusivities vs.\ temperature $T$, for fiducial
  grain parameters and three different combinations of gas density and
  plasma beta parameter.  In each panel we plot the Ohmic (solid red
  curve), ambipolar (dashed blue curve), and Hall diffusivities
  (dotted green curve; the Hall diffusivity is negative on the thicker
  portion).  Electron and ion abundances are derived using our
  chemical network including thermionic emission.  Also plotted (red
  dotted curve) is the Ohmic diffusivity derived from the Saha
  equation for potassium.  Gray shading indicates diffusivities
  greater than $h^2\Omega$, high enough that disk gas ought to be
  loaded onto the stellar fields primarily by diffusion.  The
  threshold temperature is about 1000~K and is set by ambipolar
  diffusion except in the highest density
  case.  \label{fig:magnetosphere}}
\end{center}
\end{figure}

The mass loading takes place not too far from the co-rotation radius,
where the Keplerian orbital period matches the star's rotation period.
Accreting T~Tauri stars typically rotate about once a week (Bouvier et
al.\ 1993), so the co-rotation radius around a $1 \, M_{\odot}$ star
is at $r \approx 0.07 \, {\rm AU}$.  In Figure~\ref{fig:magnetosphere}
we present results near the co-rotation point for the Ohmic
diffusivity $\eta_{\rm O}$, Hall diffusivity $\eta_{\rm H}$, and
ambipolar diffusivity $\eta_{\rm A}$, as functions of temperature.
Three combinations of density and plasma beta parameter $\beta =
P_{\rm g} / P_{\rm mag}$ are shown: $n_{\rm H2} = 10^{12} \, {\rm
  cm}^{-3}$, $\beta = 1$; $n_{\rm H2} = 10^{14} \, {\rm cm}^{-3}$,
$\beta = 1$; and $n_{\rm H2} = 10^{16} \, {\rm cm}^{-3}$, $\beta =
100$.  We choose $\beta=1$ for the first two panels, yielding a local
field strength $B$ that increases slightly with temperature.  In the
third panel we select $\beta=100$ since the stellar magnetic field at
0.07~AU is unlikely to exceed ten gauss, given a few kilogauss value
at the stellar photosphere and a dipolar falloff with distance.  The
third panel's parameters are thus more likely to be found in the
higher accretion rate regime where the inflowing gas has enough ram
pressure to force the stellar fields inward.

%-------------------------------------------------------------------------------
%
% FIGURE 20
%
%
%-------------------------------------------------------------------------------

Grey shading in Figure~\ref{fig:magnetosphere} indicates diffusivities
high enough that diffusive loading is likely to dominate.  The
transition from dynamical to diffusive loading takes place at a
temperature in the range from 960~K (at $n_{\rm H2} = 10^{14} \, {\rm
  cm}^{-3}$) to 1080~K (at $n_{\rm H2} = 10^{12} \, {\rm cm}^{-3}$).
Ambipolar diffusion causes the transition except in the case with the
highest density and weakest magnetic field, where Ohmic diffusion is
stronger. 
The Hall term is also signficant near the threshold in all three cases. 
Temperatures below the transition are quite plausible if accretion heating is weak.  
The thresholds are
comparable to the 1060~K equilibrium temperature of a blackbody
directly exposed to the starlight at 0.07~AU.  Furthermore corotation
lies beyond the dust sublimation front, so starlight can be absorbed
before reaching 0.07~AU.  In a passively heated protoplanetary disk,
temperatures at 0.07~AU are predicted to be $< 690 \, {\rm K}$
(Lesniak \& Desch 2011).  We conjecture that the stellar magnetic
field may diffuse slowly into the hot gas nearer the star, but field
lines that escape this region---either through reconnection or
disruption of the disk's inner edge by a dynamical instability---can
quickly diffuse further into the disk and load gas from the disk on
the stellar magnetosphere.

Two points seem worth emphasizing.  First, the close match between the
diffusive loading threshold and the temperatures expected in passively
heated disks means that the loading mechanism can switch depending on
conditions.  Second, at temperatures near the threshold,
thermionic and ion emission reduce the diffusivity by orders of
magnitude compared with the Saha equation.  The grains' work function
thus likely impacts the character of diffusive loading on 
young stars' magnetospheres. 
These effects should be included in future studies.

\subsection{Current Sheet Instabilities}

The treatment of high-temperature ionization in protoplanetary disks
also can be applied to the ``short-circuit'' instability of current
sheets.  In protoplanetary disks with magnetorotational turbulence, it
is reasonable to assume (and simulations show) that a commonly
occurring magnetic geometry has one region of more-or-less uniform
magnetic field adjacent to another region with the field lines
oriented in the opposite direction (Sano et al.\ 2007; Hubbard et
al.\ 2012).  The surface separating the two is marked by steep
magnetic gradients and high current densities, and is termed the
current sheet.  Ohmic dissipation can lead to heating in this sheet.
Hubbard et al.\ (2012) argued that if the gas' magnetic diffusivity
$\eta$ decreases with increasing temperature then the heating can grow
exponentially: increased temperature leads to lower diffusivity; the
current concentrates in the lower-diffusivity layer, accompanied by
more Ohmic heating; so the temperature increases further.  The runaway
heating can be quite rapid, potentially taking hours or less,
depending on the ionization properties of the gas.  Hubbard et
al.\ (2012) and McNally et al.\ (2013) further argued that this sudden
heating might be the mechanism behind the formation of chondrules,
which are known to have been melted suddenly by some ``flash-heating''
process, then cooled and crystallized over the course of hours
(Connolly et al.\ 2006).  Whether current sheet instabilities can take
place in protoplanetary disks, and whether they can grow fast enough
to be the chondrule formation mechanism, are key questions that hinge
on the ionization state of the gas at high temperatures.

The growth rate of the short-circuit instability was calculated by
Hubbard et al.\ (2012), who performed a linear analysis on a
background state in which the magnetic field increases linearly from
$-B_0/2$ to $+B_0/2$ as $x$ runs from $-L/2$ to +$L/2$, where $x$ is
the direction normal to the current sheet.  They assumed that
perturbations in the magnetic field, current density, etc., vary as
$\exp \left( +\lambda t + i k x \right)$, where $k$ is the wavenumber
of the perturbation and $\lambda$ is the growth rate.  The growth rate
is then found to satisfy the equation
\begin{equation}
\lambda^{2} + \lambda \left( t_{\rm cool}^{-1} + t_{\rm diff}^{-1} + t_{\eta}^{-1} \right) 
 +t_{\rm diff}^{-1} \, \left( t_{\rm cool}^{-1} - t_{\eta}^{-1} \right) = 0, 
\end{equation}
where $t_{\rm cool}$ is the timescale over which emission of radiation
cools the region; we specify its value below.  Here $t_{\rm diff} = 1
/ (k^2 \, {\eta_0})$ is the timescale for the magnetic field to
diffuse one wavelength, and $\eta_0$ is the magnetic diffusivity in the
background state.  Other zero subscripts likewise indicate conditions
in the background state.  The timescale $t_{\eta}$ can be defined as
\begin{equation}
t_{\eta} = -\left( \frac{ \partial \ln {\eta} }{ \partial \ln T } \right)_{T = T_0}^{-1} \, t_{\rm heat},
\end{equation}
where 
\begin{equation}
t_{\rm heat} = \frac{ 4\pi \, n_{\rm n} \, k_{\rm B} \, T_{0} }{ (\gamma-1) \, B_{0}^{2} } \, \frac{ L^2 }{{\eta}_0} 
          = \frac{ \beta_{0} }{ 2 (\gamma-1) } \, \frac{ L^2 }{{\eta}_{0}},
\end{equation}
is the timescale for Ohmic dissipation to heat the gas significantly.
We have written $t_{\rm heat}$ in terms of the plasma beta parameter
in the zero-order state, $\beta_{0} = P_{0} / (B_{0}^{2} / 8\pi)$,
which makes clear that the heating rate is directly tied to the rate
at which magnetic field lines diffuse into the current sheet.  In the
usual case where ${\eta}$ decreases with increasing temperature,
$t_{\eta} > 0$.  In that case, unstable growth is assured if $t_{\eta}
< t_{\rm cool}$.

To determine $t_{\rm cool}$ we imagine a slab of thickness $L$, heated
to temperature $T$ by Ohmic dissipation.  The slab will cool as it
emits radiation that must diffuse through the surrounding gas.  If
this gas has density $\rho$, heat capacity $C_{\rm P}$, and opacity
(per mass of gas) $\kappa$, the appropriate timescale is the radiative
diffusion time,
\begin{equation}
t_{\rm rd} = \frac{3}{64 \, \pi^2} \frac{\rho^2 \, C_{\rm P} \, \kappa}{ \sigma T^3 } \, L^2,
\end{equation}
(Mihalas \& Mihalas 1984; Kippenhahn \& Weigert 1990; Desch \&
Connolly 2002).  So that the scenario is relevant to chondrule
formation, we adopt $\rho = 1 \times 10^{-9} \, {\rm g} \, {\rm
  cm}^{-3}$, $T = 1500 \, {\rm K}$, and $C_{\rm P} = 8 \times 10^7 \,
{\rm erg} \, {\rm g}^{-1} \, {\rm K}^{-1}$.  As for the opacity, a
reasonable value would be $26 \, {\rm cm}^{2} \, {\rm g}^{-1}$ if dust
were present (at $T = 1500 \, {\rm K}$), but since micron-sized dust
generally vaporizes within minutes in the chondrule-forming region, a
more reasonable opacity would be that of chondrules alone, $\kappa
\approx 0.03 \, {\rm cm}^{2} \, {\rm g}^{-1}$ (Morris \& Desch 2010).
%
%Assuming particles with radii $300 \, \mu{\rm m}$ comprising 1\% the
%mass of the gas, we estimate $\kappa \approx 0.1 \, {\rm cm}^{2} \,
%{\rm g}^{-1}$.
% 
We find $t_{\rm rd} = 5200 \, ( L / 10^5 \, {\rm km} )^2 \, {\rm s}$
if $\kappa = 26 \, {\rm cm}^{2} \, {\rm g}^{-1}$, and $t_{\rm rd} = 6
\, ( L / 10^5 \, {\rm km} )^2 \, {\rm s}$ if $\kappa = 0.03 \, {\rm
  cm}^{2} \, {\rm g}^{-1}$.  Note that McNally et al.\ (2013) took
densities $\rho \sim 10^{-7} \, {\rm g} \, {\rm cm}^{-3}$, which would
make cooling slower by 4~orders of magnitude; but even in a disk with
10~times more material than the minimum mass solar nebula
(Weidenschilling 1977), these densities are only achieved far inside
1~AU, so some outward transport process would be needed to place the
chondrules in the asteroid belt.  At the lower densities that seem
more likely to prevail, the current sheet cools by radiative diffusion
in a $t_{\rm cool}$ of seconds to hours.

With the cooling timescale in hand, it is simple to find the ratio of $t_{\eta}$ and $t_{\rm cool}$:
\begin{equation}
\frac{t_{\eta}}{t_{\rm cool}} = -\left( \frac{\partial \ln {\eta}}{\partial \ln T} \right)^{-1} \,
 \frac{ \beta_0 }{ 2 (\gamma-1) \, {\eta}_0 } \, \frac{ 64 \pi^2 }{ 3 } \,
 \frac{ \sigma T^3 }{ \rho^2 \, C_{\rm P} \, \kappa }.
\end{equation}
Note that the ratio does not depend on the size of the slab: the heating timescale relies on diffusion of
magnetic flux, and the cooling timescale relies on diffusion of radiation, so both scale as $L^2$. 
Unstable modes require this ratio to be $< 1$, or
\begin{equation}
{\eta} > {\eta}_{\rm min} = 
\left| \frac{\partial \ln {\eta}}{\partial \ln T} \right|^{-1} \,
 \frac{ \beta_0 }{ 2 (\gamma-1) } \, \frac{ 64 \pi^2 }{ 3 } \,
 \frac{ \sigma T^3 }{ \rho^2 \, C_{\rm P} \, \kappa }.
\end{equation}
For reasonable parameters ($\rho = 10^{-9} \, {\rm g} \, {\rm cm}^{-3}$, $T = 1500 \, {\rm K}$,
$\gamma = 7/5$, $C_{\rm P} = 8 \times 10^7 \, {\rm erg} \, {\rm g}^{-1} \, {\rm K}^{-1}$), and
assuming $\beta_0 = 10$ and $\kappa = 26 \, {\rm cm}^{2} \, {\rm g}^{-1}$, 
${\eta}_{\rm min} \approx 2.4 \times 10^{17} \, \left| \partial \ln {\eta} / \partial \ln T \right|^{-1} \, 
{\rm cm}^{2} \, {\rm s}^{-1}$.
(or $\sim 10^3$ times higher if dust evaporates). 
The resistivity must be high enough that the current sheet can be continually fed with magnetic field lines 
by diffusion, faster than radiation can cool the gas.
We can also express this requirement as a {\it maximum} electron fraction $x_{\rm e}$, using the following
relationship:
\begin{equation}
{\eta} = 9060 \, \left( \frac{ T }{ 1500 \, {\rm K} } \right)^{1/2} \, x_{\rm e}^{-1} \, {\rm cm}^2 \, {\rm s}^{-1},
\end{equation}
(Hayashi 1981; Balbus \& Terquem 2001).
We find the unstable modes require
$x_{\rm e} < x_{\rm e,max} \approx 3.7 \times 10^{-14} \, \left| \partial \ln {\eta} / \partial \ln T \right|$,
or three orders of magnitude smaller if dust evaporates. 

For our canonical case depicted in Figure~2, if $T < 750 \, {\rm K}$
there is no variation in the electron density with temperature, in
which case $x_{\rm e,max} \approx 0$, and no unstable modes can be
expected.  For $T > 750 \, {\rm K}$, $\left| \partial \ln {\eta} /
\partial \ln T \right| \approx 100$ (the electron density rises 8
orders of magnitude as $T$ increases from 750~K to 900~K), so $x_{\rm
  e,max} \approx 3.7 \times 10^{-12}$ (or three orders of magnitude
smaller if dust evaporates).  This means the instability can proceed
until $T \approx 900 \, {\rm K}$.  For hotter temperatures the
electron density is too high for field lines to diffuse into the
current sheet fast enough to overcome the radiative cooling.  So, for
the short-circuit instability to act, the ambient temperature of gas
surrounding the current sheet must fall in the narrow range $750 \,
{\rm K} \, \ltsimeq \, T \ltsimeq 900 \, {\rm K}$.  The case with
grains with work function $W = 6 \, {\rm eV}$ is similar to the
canonical case.  For the $W = 2 \, {\rm eV}$ case, $\left| \partial
\ln {\eta} / \partial \ln T \right| \approx 14$ for $T$ between 600
and 800~K.  We then find $x_{\rm e,max} \approx 5 \times 10^{-13}$ or
so.  A wider range of temperatures, extending down to 650 K, is then
allowed to be unstable.  Again, if dust evaporates in the chondrule
formation region, then the faster radiative cooling means the maximum
allowable electron density is much lower, by about three orders of
magnitude.

A second requirement for the current sheet instability is that a
current sheet can be set up in the first place.  This in turn requires
the existence of magnetic turbulence in the disk, almost certainly
from the magnetorotational instability (MRI).  The MRI is suppressed
if the Ohmic Elsasser number $\Lambda_{\rm O} = (v_{\rm A}^{2} /
\Omega) / {\eta}$ is less than unity, or if ${\eta} \, \gtsimeq \,
v_{\rm A}^{2} / \Omega$, where $v_{\rm A}^{2}$ is the Alfven number
and $\Omega$ the Keplerian orbital frequency (Jin 1996).  From the
definition of the plasma beta number, $v_{\rm A}^2 = (2 / \beta) \,
C^2$, where $C$ is the sound speed; in disks in which the MRI has
saturated, typically $\beta \sim 10^{2}$ (Miller \& Stone 2000;
Hirose et al.\ 2006).  Therefore the existence of the MRI requires
${\eta} \, \ltsimeq \, 0.02 \, C^2 / \Omega$.  Note that the
existence of a current sheet demands a {\it low} value of the magnetic
diffusivity, while the short-circuit instability requires a {\it high}
value of ${\eta}$.  For reasonable parameters (say, $T \approx 800
\, {\rm K}$, at 1 AU in the disk), the MRI requires ${\eta} \,
\ltsimeq \, 2.8 \times 10^{15} \, {\rm cm}^{2} \, {\rm s}^{-1}$, or
$x_{\rm e} \, \gtsimeq \, 2.5 \times 10^{-12}$.  For our canonical
case, there is an exceedingly narrow range of electron fractions in
the ambient nebula that can both lead to current sheets and the
short-circuit instability: $2.5 \times 10^{-12} \, \ltsimeq \, x_{\rm
  e} \, \ltsimeq \, 3.8 \times 10^{-12}$, corresponding to a very
narrow temperature range about 10 K wide, centered on about 900 K.  As
the work function of solids is reduced, the range narrows.  For the $W
= 2 \, {\rm eV}$ case, there is in fact no overlap: the electron
fractions needed for runaway growth are not compatible with the
operation of the MRI and the existence of current sheets.  If dust
evaporates in the chondrule formation region, there is no overlap in
conditions for any value of $W$.  The net effect of all these factors
is that the gas ionization needed to generate magnetic turbulence and
to initiate the short-circuit instability are unlikely to be
simultaneously met in protoplanetary disks.

Finally, we point out one additional problem for the short-circuit
instability, which is its need to grow faster than the chemical
timescales will allow.  The timescales in the problem must be $<
t_{\rm cool}$, which we argue is no more than 5000~s (for $\rho =
10^{-9} \, {\rm g} \, {\rm cm}^{-3}$ and no evaporation of dust).  The
instability requires the gas resistivity to decrease as the
temperature increases, on these timescales.  The steady-state electron
abundance indeed increases with temperature, but the ionization
chemistry takes time to reach steady state.  According to Figures~2
and 5, the electron abundance starts to rise with temperature above
750~K or so, but the chemical timescales to achieve these higher
electron abundances also rise.  Above about 800~K, the chemical
timescales exceed 5000~s and approach $10^5$ s.  This effectively
reduces $\left| \partial \ln {\eta} / \partial \ln T \right|$,
potentially by 1 to 2 orders of magnitude, making the instability more
difficult to initiate and grow.

In summary, the short-circuit instability requires a rise in electron
abundance with temperature, together with an initially large magnetic
diffusivity so that magnetic heating is faster than radiative cooling.
However, generating current sheets in the first place requires a small
magnetic diffusivity.  Depending on such parameters as gas density,
opacity, grain work function, etc., there is either no way to
simultaneously satisfy these requirements, or they are satisfied only
in a narrow (perhaps 10~K) temperature range, typically above 700~K.
%We therefore cannot rule out current sheets and the short-circuit
%instability in all circumstances, but it would not seem to be a
%common occurrence in PPDs.  Moreover, the requirement that the
%temperature start above 700 K is completely inconsistent with
%existence of primary sulfur in chondrules, which requires ambient
%temperatures $< 650 \, {\rm K}$ (Lodders 2003).  Current sheets, even
%if they exist, would not seem to be the chondrule formation
%mechanism.
The short-circuit instability therefore appears to require rather
special conditions.  Moreover, the requirement that the temperature
start above 700~K is inconsistent with the presence of primary sulfur
in chondrules, which indicates ambient temperatures $<$650~K (Lodders
2003).  Even if short-circuit instabilities can arise from current
sheets, they do not seem compatible with the chondrule formation
mechanism.

\section{Conclusions}

The ionization state of the gas hotter than about 500 K
determines how well protostellar disks couple to their internal
magnetic fields and to the fields of the central star.  Properly
calculating the ionization state is especially important for assessing
where the magnetorotational instability can operate, which bears on
the question of how the disks evolve over time.  It bears on whether
disk gas can be loaded onto a star's magnetic field.  It also bears on
whether current sheets can dissipate magnetic energy via the current
sheet instability.  Previous treatments of ionization at high
temperatures have assumed that potassium atoms are in the gas phase
and that they are ionized according to the Saha equation.  This
approach ignores the role of dust and the fact that ions and electrons
strike grains far more often than they meet one another in the gas and
recombine.  Furthermore gas-phase collisions are not the only
ionization process.  At high temperatures, the grains also emit
electrons and ions into the gas.  Balancing thermionic (and ion)
emission against adsorption of electrons (and ions) on the grains
yields the Saha-Langmuir equation, which is similar in form to the
Saha equation, but instead of the first ionization potential of the
gas-phase atoms (e.g.\ 4.34~eV for potassium) involves the work
function of the grain material.  Work functions typical of the solid
materials in protoplanetary disks are near 5~eV (silicates at 5.0 to
5.4~eV, FeNi metal at 4.4~eV, graphite 4.62~eV).  These exceed the
first ionization potential of potassium, but because the grains
present so much surface area, the disk gas is more ionized (with
higher electron density) than expected based on the Saha equation.  
These processes are important at temperatures between 700 K and the
dust sublimation threshold. We are the first to identify their role 
in ionizing the gas in the central regions of protoplanetary disks.

We have incorporated thermionic and ion emission into a chemical
network and solved for the steady-state abundances of electrons and
other species as functions of temperature and density.  Unlike
previous analyses, we do not automatically assume that potassium atoms
reside in the gas phase; in fact, Lodders (2003) predicts a 50\%
condensation temperature of 1006~K, meaning that below about 1000~K,
most potassium is locked up in solids.  We mimic this effect by using
an evaporation energy $\approx 3 \, {\rm eV}$ that potassium atoms
must acquire to leave the grain surface.  Thus above 800~K, at least
one in $10^4$ or $10^5$ potassium atoms is in the gas phase rather
than locked up in the grains.  This treatment is approximate, but even
if all the sodium were firmly bound to the grains at lower
temperatures, cesium atoms are in the gas phase above 850~K (Lodders
2003) and their abundance is only $10^{4}$ times less than potassium.
Some alkali atoms thus occur in the gas below 1000~K.

The most signficant result from our canonical case is a major rise in
the electron fraction at temperatures between 700~K and 900~K.  We
find that qualitatively, the run of electron fraction with temperature
is relatively insensitive to the work function (Figure~8), gas density
(Figure~16), and dust-to-gas mass ratio (Figure~12).  Quantitatively,
the temperature at which the electron fraction rises sharply does vary
somewhat: if the dust-to-gas mass ratio is increased (decreased) two
orders of magnitude, the rise in electron density is shifted about
100~K to higher (lower) temperatures (Figure~12), with similar results
expected for changes in grain size.  Likewise, increasing (decreasing)
the gas density by two orders of mangitude causes the rise in electron
density to occur at temperatures about 100~K higher (lower)
(Figure~16).  A notable feature of the results is that the electron
density can {\it almost} be computed using the Saha equation for
potassium, but one must use an ionization potential less than 4.34~eV.
Above 800~K, for plausible grain work functions, values between 3.5
and 4.0~eV yield the correct ionization state, and that state is
significantly more ionized than at lower temperatures.  As a robust
result, then, thermal ionization can be said to be significant above
about 800~K.

A notable feature of the high-temperature regime is that the grains
become rather highly charged through directly emitting electrons (or
ions).  The grain charge can be positive or negative, but for the most
likely work function values near 5~eV, the grains emit ions and are
left with negative charges.  The charging is strong above 800~K, and
grains of radius $1 \, \mu{\rm m}$ reach charges of about $-700 \, e$
at 900~K (Figure~3).  While much larger than the grain charges found
neglecting thermionic and ion emission, this is equivalent to an
electric potential of just 1~volt.  The charging is strong enough to
affect the grains' coagulation at high temperatures.  The Coulomb
energies that must be overcome for grains to collide are $\sim
10^3$~eV, or $\sim 10^{-9} \, {\rm erg}$, so micron-sized grains
coming together slower than $10 \, {\rm cm} \, {\rm s}^{-1}$ will be
mutually repelled by the electrical force.  The effects of grain
charging on the coagulation of grains have been investigated by
Okuzumi (2009).  We also point out that if there were any grains
composed of materials with work functions $\ltsimeq \, 3.5$~eV, then
their positive charges would enhance their collision rates with the
negatively charged grains.

We have applied our charging model to the question of the MRI dead
zone's inner edge.  At low temperatures the gas is insufficiently
ionized to permit the coupling of the gas to the magnetic field that
is fundamental to the MRI.  We find that for reasonable disk
parameters (gas density, magnetic field strength), the temperature
needed for the MRI to operate is set by ambipolar rather than Ohmic
diffusion.  The typical threshold temperatures are in the range 900 to
1100~K.
At the high densities found near the midplane, the thresholds are 
comparable to or up to 100 K less than when neglecting the thermionic
and ion emission and considering only the Ohmic resistivity.
At the lower, atmospheric densities, ambipolar diffusion makes the 
thresholds up to 300 K higher than those from the Saha equation and
Ohmic resistivity.
Nevertheless, the results validate 
the common belief that thermal ionization requires
temperatures above 1000~K.  We likewise have investigated the
temperatures required to diffusively load the disk mass onto the
stellar magnetosphere.  We find once again that temperatures below
1000~K are required, and that the threshold is set by ambipolar
diffusion.

The high-temperature ionization also has implications for the onset of
the ``short-circuit instability'' and its role in chondrule formation.
This instability relies on the resistivity of the gas decreasing with
increasing temperature, so that dissipation of magnetic energy during
reconnection events can accelerate.  We confirm that above a critical
threshold temperature (about 800 K in our canonical case, but
potentially as low as 650 K), this condition is met.  
On the other hand, the presence of primary sulfur (Rubin et al.\ 1999)
in chondrules suggests 
that they condensed in regions cooler than 650 K, 
the condensation temperature of sulfur. 
In addition, we show that there are two  extra 
and seemingly incompatible
constraints with which the short-circuit instability must comply.
The instability requires the resistivity to change on timescales
faster than the gas can cool.  This in turn imposes a {\it lower}
limit on the resistivity, and an upper limit on the electron fraction,
so that field lines can diffuse rapidly enough into the reconnection
region; typically $x_{\rm e}$ must not exceed about $10^{-13}$.  The
short-circuit instability also requires magnetic turbulence to set up
the current sheets of oppositely oriented field lines.  Assuming
magnetic turbulence is generated by the MRI, this in turn imposes an
{\it upper} limit on the resistivity, and a lower limit to the
electron fraction; typically $x_{\rm e}$ must be $10^{-13}$ or
greater.  For typical conditions, only a small range of electron
fractions satisfy both conditions simultaneously, while for many
parameters (especially if dust evaporates in the chondrule-forming
region), no conditions simultaneously lead to the formation of current
sheets and allow the short-circuit instability.  
This and the constraint that chondrule precursors started at temperatures $< 650 \, {\rm K}$
to retain primary sulfur would seem to imply that current sheets and the
short-circuit instability are probably not the chondrule formation
mechanism.

We note that thermionic and ion emission may be important to the
ionization state of gas in the atmospheres of exoplanets, especially
``hot Jupiters''.  Ohmic dissipation of magnetic energy in their
atmospheres has been invoked as the explanation for the inflated radii
(above model predictions) of hot Jupiters such as HD209458b and
HD189733b (Batygin \& Stevenson 2010; Perna et al.\ 2010) and may
affect atmospheric circulation (Menou 2012; Rogers \& Komacek 2014).
At present, the resistivity of the gas in such models is computed
essentially using equation~41 (as in Balbus \& Terquem 2001), with the
electron density found using the Saha equation for various gas-phase
atoms (Batygin \& Stevenson 2010; Rogers \& Komacek 2014).  However,
hot Jupiters, HD209458b and HD189733b in particular, are predicted to
have cloud layers containing condensates such as TiO (Fortney et
al.\ 2010).  TiO has a work function of about 4.5~eV (Fomenko 1966;
Greiner et al.\ 2012) and a cosmochemical abundance equivalent to a
dust-to-gas ratio $\sim 10^{-5}$ (Lodders 2003).  At pressures of 1
bar, and temperatures of 1000 K, the gas density would be $n_{{\rm
    H}_{2}} \, \approx \, 6 \times 10^{18} \, {\rm cm}^{-3}$.  We
therefore expect that thermionic and ion emission from TiO particles,
and adsorption of electrons and ions onto them, will control the
ionization state of the gas, which will then {\it not} be described
using a Saha equation, above some threshold temperature.  Comparing to
Figures~12 and 16, we expect that the high density will raise the
threshold temperature, while the low dust-to-gas ratio would lower the
threshold temperature, but that thermionic emission would dominate the
charging of the gas above about 800 to 1000 K.  Naively, we expect
thermionic emission will lead to higher electron fractions and lower
Ohmic dissipation than use of the Saha equation alone, but exoplanet
atmospheres are far enough out of the regime considered in this paper
that other details may play a role.  In any case, we recommend
that this charging mechanism be considered when calculating Ohmic
dissipation in the atmospheres of hot Jupiters.

We also recommend that future effort be directed to measuring the work
functions of astrophysically relevant materials, especially those
found in protoplanetary disks and exoplanet atmospheres.  We culled
information from a variety of decades-old sources, but a dedicated
effort should be made to consistently compare thermionic and ion
emission from silicates, FeNi, graphite particles, and other
astrophysical solids.

Where solids exist, they can catalyze the recombinations of electrons
and ions on their surfaces, rendering the Saha equation incorrect.
But they also emit electrons or ions directly into the gas, at rates
that depend on their material properties.  In the end, the ionization
fraction increases faster with temperature than under the Saha
equation.  For relevant conditions, we find that the ionization rises
significantly above 800~K.

\acknowledgments We thank the organizers of the August 2014 meeting
``Non-ideal magnetohydrodynamics, stability, and dissipation in
protoplanetary disks'' at the Niels Bohr International Academy in
Copenhagen, for providing a forum for our collaboration.  This
research was carried out in part at the Jet Propulsion Laboratory,
California Institute of Technology, under a contract with the National
Aeronautics and Space Administration, and with the support of grant
number 13-OSS13-0114 from the Solar Systems Origins program.

%% in parentheses, after the keyword, but they will not be verified.

%{\it Facilities:} \facility{Nickel}, \facility{HST (STIS)}, \facility{CXO (ASIS)}.



%% In this first example, note that the \tabletypesize{}
%% command has been used to reduce the font size of the table.
%% We also use the \rotate command to rotate the table to
%% landscape orientation since it is very wide even at the
%% Note also that the \label command needs to be placed

%% This table also includes a table comment indicating that the full
%% version will be available in machine-readable format in the electronic

\end{document}